%% file: main.tex
\documentclass[conference]{IEEEtran}

\usepackage{amsmath,amssymb}
\usepackage{booktabs}
\usepackage{array}
\usepackage{graphicx}
\usepackage[letterpaper,left=1in,right=1in,top=1in,bottom=1in]{geometry}
\usepackage[hidelinks]{hyperref}
\usepackage{placeins}
\usepackage{longtable}
\usepackage{pdflscape}

\renewcommand{\small}{\normalsize}

\renewcommand{\scriptsize}{\normalsize}

\newcommand{\tool}{VesperCSI}

\title{VesperCSI: A High-Performance and Extensible CSI Acquisition Framework for Edge Wi-Fi Sensing}
\author{
\IEEEauthorblockN{Kaan Baydemir}
\IEEEauthorblockA{REAL LAB\\
EE Department, METU\\
Ankara, Turkey\\
kaan.baydemir@metu.edu.tr}
\and
\IEEEauthorblockN{Sajjad Baghaee}
\IEEEauthorblockA{REAL LAB\\
EE Department, METU\\
Ankara, Turkey\\
sajjad@baghaee.com}
\and
\IEEEauthorblockN{Ilkay Ulusoy}
\IEEEauthorblockA{REAL LAB\\
EE Department, METU\\
Ankara, Turkey\\
ilkay@metu.edu.tr}
}

\begin{document}
\maketitle

\begin{abstract}
Channel State Information (CSI) captures fine-grained changes in Wi-Fi channel
propagation and provides a foundation for device-free sensing applications such
as activity recognition, localization, and occupancy detection. The development of such applications depends on an acquisition
system that can handle high-rate CSI streams while remaining
extensible to the requirements of individual experiments. \tool{} is an
open-source, GPL-3.0-licensed redesign of FeitCSI for CSI acquisition with
Intel AX200 and AX210 Wi-Fi interfaces. It targets high-rate acquisition on
amd64 and aarch64 Linux hosts using fixed-capacity preallocated object pools
and separate capture and consumer threads. A loadable plugin interface
supports experiment-specific monitoring, storage, and processing. We evaluate
\tool{} and FeitCSI on Raspberry Pi CM4 receivers in a burst-traffic software
crossover. At a 600-MHz all-core frequency ceiling and a target rate of
2500~frames/s, \tool{} achieves mean sequence retention of
99.017--99.087\% in 10,000-frame bursts, compared with 43.972--46.553\% for
FeitCSI. This ordering persists after exchanging the userspace programs
between two receivers while retaining their respective drivers. The comparison
therefore characterizes the tested software-and-driver packages rather than
isolating userspace effects. These results support \tool{} as a reusable
acquisition component for edge-hosted Wi-Fi sensing under the evaluated
workload and platform conditions.

\end{abstract}


\section{Introduction}

Wi-Fi Channel State Information characterizes the complex response of
individual OFDM subcarriers and antenna paths. Its sensitivity to propagation
changes has enabled device-free activity recognition, gesture sensing,
localization, and occupancy estimation. The Linux
802.11n CSI Tool established practical CSI access on commodity Intel hardware
\cite{halperin2011}. Subsequent platforms extended extraction to Atheros NICs
\cite{atheros}, Broadcom devices \cite{nexmon}, commercial 802.11ax hardware
\cite{axcsi}, and Intel AX200/AX210 interfaces \cite{feitcsi,picoscenes}.

Although these developments have expanded CSI accessibility, acquisition on a
compact host remains a systems problem. Bursty netlink input must be processed
without per-record allocation, malformed or irrelevant records must be rejected
before storage, and output or analysis work must not unnecessarily obstruct
ingestion.

\tool{} addresses this gap through a high-performance, extensible acquisition
program implemented in C. It supports amd64 and aarch64 Linux hosts, including
Raspberry Pi-class systems equipped with a compatible PCIe Intel AX200 or AX210
interface. CSI access is provided through a CSI-enabled driver build based on Intel's
maintained \texttt{backport-iwlwifi} tree rather than through the driver
distributed with the Linux kernel tree. VesperCSI supplies the userspace acquisition engine,
configuration lifecycle, storage path, and experiment-control workflow above
this driver boundary.

In two preceding REAL Lab studies, we used FeitCSI to acquire CSI for an
evaluation of preprocessing methods and for a dataset used to benchmark
deep-learning architectures for Wi-Fi human activity recognition
\cite{celik2026preprocessing,emiroglu2026realcsi}. Experience gained while
conducting these studies, together with the requirement to move future
experiments onto aarch64 edge hosts equipped with commodity Intel AX200/AX210
interfaces, motivated the development of a reusable acquisition tool tailored
to this deployment setting.


\setlength{\tabcolsep}{1pt}
\renewcommand{\arraystretch}{1.05}
\begin{table*}
\centering
\caption{Comparison of reusable CSI acquisition platforms from public
documentation and repositories. BW denotes the maximum documented CSI channel
bandwidth. ND denotes ``not documented,'' and N/A denotes a non-comparable
execution model.}
\label{tab:comparison}
\resizebox{\textwidth}{!}{\begin{tabular}{@{}>{\raggedright\arraybackslash}p{2.0cm} >{\raggedright\arraybackslash}p{3.3cm} >{\raggedright\arraybackslash}p{3.6cm} >{\raggedright\arraybackslash}p{2.5cm} >{\centering\arraybackslash}p{2.0cm} >{\raggedright\arraybackslash}p{2.6cm} >{\centering\arraybackslash}p{1.2cm} >{\centering\arraybackslash}p{1.3cm} >{\raggedright\arraybackslash}p{3.6cm}@{}}
\toprule
\textbf{Tool} & \textbf{Source and license} &
\textbf{Plugins or custom workflows} & \textbf{Concurrency} &
\textbf{BW} & \textbf{Acquisition hardware} &
\textbf{amd64} & \textbf{aarch64} & \textbf{Deployment notes} \\
\midrule
Linux 802.11n CSI Tool \cite{halperin2011,intelcsitool} &
Open source with mixed terms: MIT userspace, GPLv2 kernel modifications, and
separate Intel firmware terms & Scripts and modifiable source; no documented runtime
plugin API & Serial reference logger & 40 MHz & Intel 5300 & ND & ND & Legacy Ubuntu/Linux
environment and modified driver/firmware; MATLAB-oriented analysis \\

Atheros CSI Tool \cite{atheros,atherosapp} & GPLv2 kernel; public CSI logger
source, license ND & Source and userspace
utilities; no documented runtime plugin API & Serial reference logger & 40 MHz & Supported Atheros 802.11n devices using
the modified \texttt{ath9k} path & ND & ND & Ubuntu PCs and OpenWrt-class
routers; additional embedded architectures are hardware-specific \\

Nexmon CSI \cite{nexmon,nexmonrepo,nexmonlicense} & Public source; custom citation
and military-use restrictions; GPL-3.0 base framework & Firmware-patch framework, extraction configuration, and
standard packet-capture tools; no documented host plugin API & ND & 80 MHz & Selected
Broadcom/Cypress Wi-Fi chips & Build & Yes & Firmware-level extraction
on Android devices, routers, and Raspberry Pi 3B+/4B/5 \\

AX-CSI \cite{axcsi,axcsiproject} & Research source archive available on
request; license not stated & Driver/firmware research release; no documented
plugin API & ND & 160 MHz & Broadcom BCM43684 802.11ax access point & N/A & ND &
Modified firmware/driver on Asus RT-AX86U; reports sent to a host over UDP \\

FeitCSI \cite{feitcsi} & Open source; GPL-3.0 & CLI, GUI, and modifiable source;
no documented plugin API & Yes; receive worker and synchronized plot queue
& 160 MHz & Intel AX200/AX210 & Yes & Yes (this work) & Linux extraction and
injection for 802.11a/g/n/ac/ax; project documentation claims amd64 and ARM
availability \\

PicoScenes \cite{picoscenes,picoscenesplugin,picoscenespdk,picoscenesfeatures} & Licensed core;
open PDK components under GPL-3.0 & Multiple concurrently active runtime plugins,
user commands, APIs, and measurement programs & Yes; multithreaded SDR reception/decoding
& 160 MHz COTS; 320 MHz SDR & Intel AX and other NICs plus SDRs & Yes &
No & Broad experimental scope; installation requirements specify amd64
and explicitly exclude aarch64 \cite{picoscenesinstall} \\

IAX \cite{iax,iaxrepo} & MIT project code; GPLv2 driver and separate firmware
terms & Operating modes, scripts, and TCP
streaming support custom pipelines; no documented runtime plugin API & Serial C++ listener & 160 MHz & Intel
AX200/210/211 and Intel 5300; AX201 provisional & ND & ND & Ubuntu 20.04/22.04 with
Linux 5.15.x; AX201 firmware absent and untested in the repository \\

ESP-CSI \cite{espcsi} & Open source; Apache-2.0 & ESP-IDF components and examples
can be composed at build time; no documented host runtime plugin loader & RTOS task model &
40 MHz (chip-dependent) & ESP32-family microcontrollers & N/A & N/A & Native Xtensa or RISC-V
embedded deployment; host computers are used only for control and analysis \\

ZTECSITool \cite{ztecsi,ztecsirepo} & Public source; no repository license
identified & Python GUI and command protocol are modifiable; no documented
plugin API & Yes; host listener/display threads; no capture queue
& 160 MHz & ZTE E2631 / SR6110
access points & ND & ND & CSI-enabled AP firmware; a separate Python host
configures reporting and receives UDP reports \\

openwifi CSI \cite{openwifi,openwificsi} & Open source; AGPL-3.0 with GPL/BSD/ISC components &
Complete HDL, driver, and software stack supports source-level customization;
no documented runtime plugin API & FPGA extraction; userspace polling & 20 MHz & Xilinx
Zynq/ZynqMP SDR/FPGA boards & N/A & Yes (ZynqMP) & Custom FPGA image and Linux
driver; prioritizes PHY programmability over commodity-NIC portability \\

\tool{} \cite{vespercsi} & Open source; GPL-3.0 & Runtime plugin interface with one active plugin
per process; bundled monitoring and experiment-control workflows & Yes; separate
capture and consumer threads & 160 MHz & Intel AX200/AX210 & Yes & Yes & Linux
with a supported PCIe NIC and CSI-enabled \texttt{backport-iwlwifi}; tested on
Linux 6.8 and 6.12, including Debian 13, and on Raspberry Pi CM4
platforms \\
\bottomrule
\end{tabular}}
\end{table*}
\setlength{\tabcolsep}{6pt}
\renewcommand{\arraystretch}{1.0}

These requirements led to \tool{}, which was developed as a hard fork of
FeitCSI, an Intel AX200/AX210 CSI acquisition platform \cite{feitcsi}. FeitCSI
provides the project's software foundation, while \tool{} substantially
redesigns the userspace internals around preallocated object pools,
consumer-driven pool exchange, explicit overflow policies, and a loadable
plugin interface intended for experiment-specific workflows. The contributions
presented here concern these architectural changes and workflow extensions
above the Intel CSI extraction path, with explicit acknowledgment of their
FeitCSI origin. \tool{} is distributed under GPL-3.0 \cite{vespercsi}.

The target deployment combines an aarch64 edge host, an Intel AX200/AX210
receiver, and a Linux 6.x kernel. Within this scope, this paper makes three
contributions: (1) a redesigned, bounded-buffer acquisition pipeline that
achieves higher sequence retention than the tested FeitCSI package under the
evaluated burst-traffic conditions; (2) a versioned plugin interface,
demonstrated through an experiment-specific activity-labeling workflow, that
supports extensible monitoring, storage, and processing; and (3) corrections
to implementation bugs inherited from FeitCSI. The retention comparison is
package-level evidence and does not isolate the userspace changes from the
respective drivers. The target platform also motivates the selection of a
directly comparable acquisition baseline in Section~\ref{sec:edge-positioning}.

The remainder of the paper is organized as follows. Section~II surveys existing
CSI acquisition platforms and compares their capabilities and deployment
requirements. Section~III describes the AX200/AX210 firmware and CSI-enabled
driver requirements for the target hardware. Section~IV details VesperCSI's
capture architecture, bounded-buffer pipeline, record validation, and
acquisition lifecycle. Section~V presents the versioned plugin interface and
demonstrates it with an infrared-controlled activity-labeling workflow.
Section~VI positions VesperCSI for ARM edge acquisition and motivates the
choice of comparison baseline. Section~VII describes the experimental setup
and evaluates burst retention, FeitCSI memory growth, and sustained VesperCSI
acquisition. Section~VIII discusses the study's limitations and directions
for future work. Section~IX summarizes the system, principal findings, and
scope of the evidence.

\section{CSI Tool Landscape}

CSI software spans acquisition, file parsing, signal processing, and sensing
applications. This section focuses on reusable acquisition platforms:
systems that obtain CSI from a NIC, access point, embedded radio, or SDR and
make the measurements available to a host application. Parsers such as CSIKit
and csiread remain useful downstream tools, but are excluded from the table
because they do not configure a radio or acquire CSI directly
\cite{csikit,csiread}.

Table~\ref{tab:comparison} presents a feature and deployment comparison of
these acquisition platforms, rather than a head-to-head performance ranking.
It focuses on source and license availability, extension mechanisms,
acquisition concurrency, documented channel bandwidth, supported hardware and
host architectures, and practical deployment constraints. The entries are
based on project papers, documentation, and public repositories reviewed on
October 8, 2026. ``Open source'' is used only when the relevant implementation
is distributed under an identifiable open-source license; a public repository
without a license is reported as public source. ``ND'' means that the
capability was not documented or could not be established from the reviewed
materials, not that a private or experimental implementation does not exist.
Architecture entries refer to the acquisition deployment rather than merely
to a machine that can compile utilities or analyze resulting files; ``Build''
denotes build-host support only. Concurrency entries describe the referenced
acquisition program or extraction path, not every component of the platform.
Likewise, the absence of a documented plugin API does not rule out
source-level extensions.

\subsection{Source Availability and Extensibility}

The reviewed projects expose several different forms of openness. FeitCSI,
ESP-CSI, and openwifi publish source under GPL-3.0, Apache-2.0, and AGPL-3.0,
respectively, with separate terms for bundled dependencies
\cite{feitcsi,espcsi,openwifi}. IAX's MIT project license coexists with GPLv2
driver code and separate firmware terms. The Atheros kernel is GPLv2, but a
license for its separate CSI logger was not identified
\cite{atheros,atherosapp,iaxrepo}. The Linux 802.11n CSI Tool likewise has
separate userspace, kernel, and Intel firmware terms \cite{intelcsitool}.
Nexmon CSI contains custom file-level terms requiring citation and restricting
military-related use, alongside the GPL-3.0 Nexmon framework \cite{nexmonlicense}.
PicoScenes retains a licensed core and publishes a GPL-3.0 plugin development
kit; ZTECSITool's repository has no identified license
\cite{picoscenespdk,ztecsirepo}. AX-CSI's research archive is request-based with
unstated reuse terms \cite{axcsiproject}. \tool{} provides its source under
GPL-3.0 \cite{vespercsi}.

Extensibility must likewise be separated into runtime composition and source
modification. PicoScenes has the most general documented runtime composition
model: it discovers multiple plugins and can deliver measurements to multiple
active plugin instances \cite{picoscenesplugin}. \tool{} also exposes a runtime
plugin boundary, but currently loads exactly one shared object per process.
The selected plugin is initialized at startup and runs in the consumer path for
each accepted record. The interface is intended to support experiment-specific
monitoring, data handling, and control workflows. Only one plugin can currently
be active within a process.

Other tools permit useful custom workflows without an equivalent loader.
FeitCSI and IAX expose command-line modes and modifiable source; Nexmon CSI uses
a firmware-patching and packet-capture workflow; ZTECSITool provides a separate
Python host GUI that configures AP-side CSI extraction and receives UDP reports
\cite{ztecsirepo}; ESP-CSI composes ESP-IDF components at build time; and
openwifi exposes the HDL, driver, and software stack. FeitCSI additionally
provides a UDP-socket mode for interaction with external tools such as MATLAB
\cite{feitcsi}. This enables out-of-process processing, but CSI delivered to an
external consumer must traverse the socket interface. In contrast, \tool{}
plugin callbacks operate directly on consumer-local records within the
acquisition process, avoiding a mandatory socket hop for extensions that can
run in process. This distinction may reduce data movement and protocol-processing
overhead, but its performance impact was not measured in this work.

\subsection{Concurrency and High-Rate Acquisition}

Concurrency is inconsistently documented across CSI projects, so the table
does not infer multithreading from implementation language or advertised data
rate. The Intel 5300 and Atheros reference loggers and IAX's C++ listener
receive and write records in serial loops. FeitCSI uses worker threads and a
synchronized plotting queue, while file output runs in the receive callback
\cite{feitcsi,intelcsitool,atherosapp,iaxrepo}. PicoScenes explicitly documents
multithreaded SDR reception and decoding \cite{picoscenesfeatures}.
ZTECSITool has host listener/display threads, with CSV writing in the receive
thread and plotting of the latest sample \cite{ztecsirepo}.
ESP-CSI uses ESP-IDF tasks; openwifi combines FPGA extraction with userspace
polling \cite{espcsi,openwificsi}.

\tool{} makes the concurrency boundary explicit. A capture thread removes CSI
messages from the kernel-facing receive path and transfers accepted records
through preallocated pools. A consumer thread then performs storage and plugin
execution. This separation prevents ordinary file or plugin work from executing
in the reception thread, while the fixed pool capacities make buffering and
overflow behavior bounded. The architecture is therefore described as
multithreaded for acquisition, not merely because auxiliary interface threads
exist.

\subsection{Bandwidth and Radio Generation}

The Linux 802.11n and Atheros tools are limited to 20/40-MHz 802.11n-era
measurements, while Nexmon CSI documents extraction up to 80 MHz on supported
Broadcom devices \cite{atheros,intelcsitool,nexmonrepo}. AX-CSI, FeitCSI, IAX,
ZTECSITool, PicoScenes on supported commodity NICs, and \tool{} reach 160 MHz,
which is the maximum channel width of the Intel AX210 generation and a
useful boundary for contemporary commodity 802.11ax CSI acquisition
\cite{axcsi,feitcsi,picoscenes,iax,ztecsi,intelax200,intelax210}. PicoScenes
also documents 320-MHz SDR-based CSI measurement on suitable hardware
\cite{picoscenesfeatures}, but that capability should not be
interpreted as 320-MHz support by the Intel AX NICs. ESP-CSI remains an
embedded, low-cost alternative at up to 40 MHz, whereas openwifi emphasizes an
open and programmable PHY rather than wide commodity channels.

The 160-MHz platforms are not interchangeable. AX-CSI and ZTECSITool instrument
specific access-point hardware; PicoScenes spans multiple commercial NICs and
SDRs; and FeitCSI, IAX, and \tool{} use Intel AX-series client interfaces.
Within the latter group, AX200 covers 2.4/5-GHz operation and AX210 adds 6-GHz
Wi-Fi 6E capability. Actual 6-GHz use also depends on the driver, firmware,
access point, and regulatory configuration.

\subsection{Host Architecture and Software Baseline}

Architecture portability depends on both userspace code and the radio-facing
driver or firmware. Older Intel and Atheros platforms document conventional PC
or router environments but do not establish current aarch64 support. Nexmon
CSI and ESP-CSI run close to their target Broadcom and ESP32 radios, while
openwifi documents 64-bit acquisition on ZynqMP boards \cite{openwificsi}.
PicoScenes requires amd64
and explicitly excludes ARM \cite{picoscenesinstall}. IAX documents
Ubuntu 20.04/22.04 and a Linux
5.15.x baseline, but its repository does not claim an aarch64 deployment
\cite{iaxrepo}. FeitCSI advertises amd64 and aarch64 availability
\cite{feitcsi}; its aarch64 execution is demonstrated by the CM4 evaluation
in this work. Generic PC or ARM support alone does not establish a particular
64-bit architecture.

\tool{} supports both amd64 and aarch64 Linux hosts, provided that the system
can attach a compatible PCIe Intel AX200/AX210 interface and build the required
CSI-enabled \texttt{backport-iwlwifi} driver. It has been verified with Linux
6.8 and 6.12, including Debian 13 deployments, and on Raspberry Pi Compute
Module 4 platforms. These kernel versions are tested
configurations rather than compatibility bounds. Operation on an older or newer
kernel depends on whether the required \texttt{backport-iwlwifi} source can be
compiled and loaded for that target kernel. This combination of two host
architectures, 160-MHz Intel AX acquisition, an explicit two-thread data path,
and a loadable workflow boundary distinguishes \tool{} from source-modifiable
loggers that provide no runtime extension interface. The driver and firmware
requirements underlying this portability are discussed in the following
section.

\section{Intel AX200/AX210 Firmware and Driver Requirements}

The firmware and host driver have distinct roles in Intel AX-series operation.
The AX200 and AX210 execute Intel microcode that is loaded by the Linux driver
during device initialization. VesperCSI does not require custom replacement
firmware: the compatible Intel firmware already supplied by the target
distribution may be retained. If the required files are absent, they can be
obtained through the official \texttt{linux-firmware} distribution and placed
in the distribution's firmware directory, commonly \texttt{/lib/firmware}
\cite{iwlwifi}.

The standard distribution-provided \texttt{iwlwifi} driver, however, does not
provide the CSI delivery interface required by VesperCSI. CSI acquisition
therefore requires the target system's wireless driver stack to be replaced
with a CSI-enabled build based on Intel's maintained
\texttt{iwlwifi/backport-iwlwifi.git} repository. This backport tree contains
the relevant \texttt{iwlwifi}, \texttt{mac80211}, and \texttt{cfg80211}
components and is retrieved from the Intel Linux Wireless repository hosted on
\texttt{git.kernel.org} \cite{backportiwlwifi}. The resulting modules must be
built for and installed on the target kernel, then loaded in place of the
distribution's original wireless modules before VesperCSI is started.

Driver installation is consequently a required system-preparation step, while
firmware replacement is not. The selected backport branch or commit must also
be compatible with the target kernel; the Linux Wireless documentation notes
that older backport revisions may not build or operate against newer kernels
\cite{iwlwifi}. For reproducibility, a VesperCSI deployment should record the
kernel version, firmware revision, and exact \texttt{backport-iwlwifi} commit
used for acquisition.

\section{VesperCSI Implementation}

\subsection{System Architecture and Bounded Capture Pipeline}

\tool{} incorporates radio configuration and process lifecycle management into
the acquisition path. At startup, it identifies the selected wireless PHY,
records and removes the existing interface, creates a monitor interface,
applies the requested channel configuration, and enables CSI through the
installed CSI-enabled \texttt{backport-iwlwifi} driver. It starts the consumer
worker before the capture worker. A shared shutdown flag and a pollable pipe
coordinate termination with the main thread and the kernel-facing receive
loop. The pool, synchronization, and shutdown mechanisms described below
separate reception from consumer work without implying lossless capture
\cite{vespercsi}.

\begin{figure*}[!t]
\centering
\includegraphics[width=\textwidth]{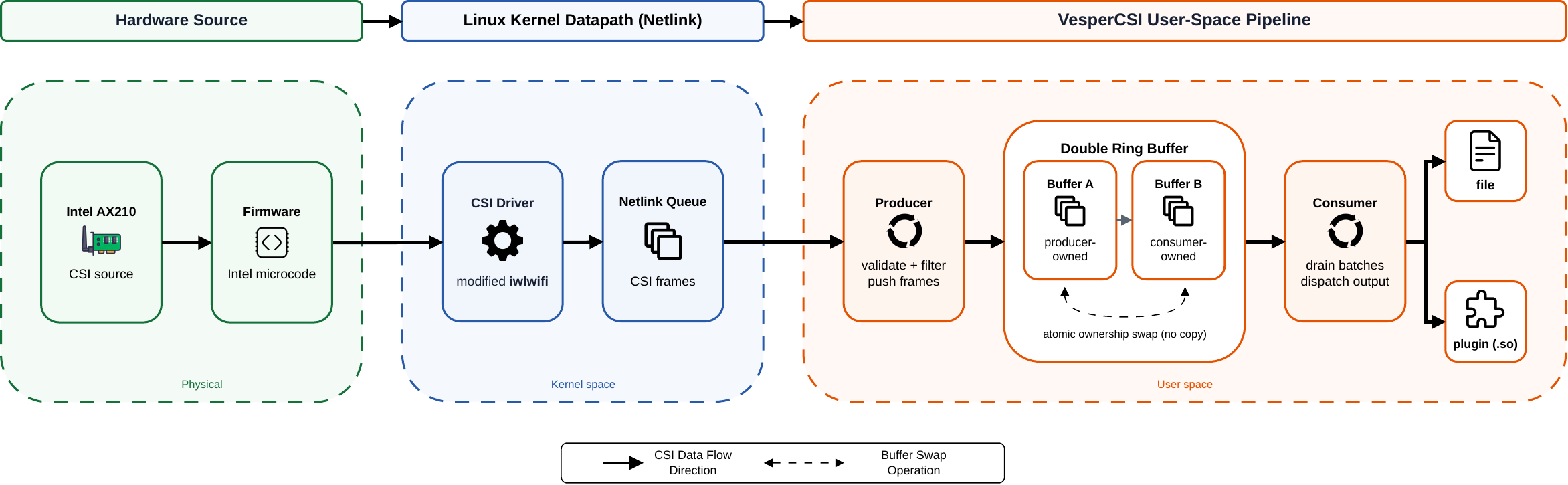}
\caption{End-to-end \tool{} dataflow from the AX-series hardware and CSI-enabled
driver to the userspace producer, double-buffer ownership exchange, consumer,
and plugin and/or file output. The ownership swap transfers pool roles without
copying queued records between pools; enqueue and dequeue operations still
copy individual CSI objects.}
\label{fig:dataflow}
\end{figure*}

Fig.~\ref{fig:dataflow} presents the complete CSI dataflow. The Intel AX210
shown in the figure represents the supported AX-series source; the same
software path is used with an AX200. CSI measurements produced by the NIC and
its standard Intel firmware enter the CSI-enabled \texttt{backport-iwlwifi}
driver, which delivers the frames to a kernel netlink queue. The VesperCSI
producer receives frames from this queue, validates their structure, applies
the configured filters, and pushes accepted records into the producer-owned pool.

\subsubsection{Pool Layout and Memory Estimate}
\label{sec:pool-layout}

The acquisition state contains two identically sized, statically reserved
pools. A rough estimate of their memory footprint follows from the number
of record slots and the size of the CSI objects stored in them. Memory
sizes below are expressed in bytes and pool capacity as a number of records.

The capacity of each pool is determined by two compile-time parameters.
The nominal record rate used for sizing is denoted by $R_{\max}$ and
expressed in records/s. The corresponding nominal buffering window is
denoted by $T_{\max}$ and expressed in seconds.
Their product defines $N$, the number of nodes and maximum queued-record
count in one pool:
\begin{equation}
    N = R_{\max} \times T_{\max}.
\end{equation}
The two pools therefore contain $2 \times N$ nodes in total. These parameters size
the arrays; they do not enforce a runtime rate limit or trigger a timed
handoff. $R_{\max}$ is not a measured sustainable rate, and $T_{\max}$ is
not a guaranteed loss-free buffering duration.

Each reserved node stores a CSI object containing a header and fixed-capacity
data storage. Let $S_{\max}$ denote the maximum supported number of subcarriers.
Let $A_{\mathrm{tx}}$ and $A_{\mathrm{rx}}$ denote the maximum numbers of
supported transmit and receive antenna paths, respectively. The data array therefore
reserves space for $S_{\max} \times A_{\mathrm{tx}} \times A_{\mathrm{rx}}$
subcarrier--path
combinations. Each complex CSI coefficient is represented by a signed 16-bit
in-phase (I) value and a signed 16-bit quadrature (Q) value, for a total of
4 bytes per combination. The maximum data-array size in bytes, denoted by $D$,
is therefore
\begin{equation}
    D = 4 \times S_{\max} \times A_{\mathrm{tx}} \times A_{\mathrm{rx}}.
\end{equation}
With $H$ denoting the header length in bytes, the estimated storage per CSI
record, $C$, is
\begin{equation}
    C = H + D
      = H + (4 \times S_{\max} \times A_{\mathrm{tx}} \times A_{\mathrm{rx}}).
\end{equation}
This sizing model excludes pointers, other bookkeeping fields, and alignment
padding; $C$ is therefore not an exact structure-size calculation. It describes
the reserved header and data storage, rather than the variable serialized
length of a particular measurement. A shorter payload does not shrink a
reserved node.

Since each pool reserves $N$ record slots, its estimated CSI storage footprint
$P$ follows by multiplying the per-record estimate by the pool capacity:
\begin{equation}
    P = N \times C = R_{\max} \times T_{\max} \times C.
\end{equation}
The total estimated CSI storage across both pools, $P_{\mathrm{total}}$, is
consequently
\begin{equation}
\begin{aligned}
    P_{\mathrm{total}} &= 2P
      = 2R_{\max}T_{\max}C\\
    &= 2R_{\max}T_{\max}
       \bigl(H + 4S_{\max}A_{\mathrm{tx}}A_{\mathrm{rx}}\bigr).
\end{aligned}
\end{equation}
Both $P$ and $P_{\mathrm{total}}$ are expressed in bytes. The actual pool
allocation additionally includes the omitted record and queue-management
overhead, including linked-list pointers, pool metadata, and padding.
Initialization links unused
nodes into a free list, and enqueue and dequeue reuse the reserved storage
without per-record heap allocation by the pool. 

\subsubsection{Consumer-Driven Ownership Exchange}

The producer inserts accepted records at the front of its pool; the consumer
removes records from the back of its own pool, giving FIFO service within
each batch. Insertion copies the CSI object into a free node. Removal copies
it into a consumer-local object and returns the node to that pool's free
list before invoking output processing. Thus, the handoff avoids copying
an entire batch between pools, but the record path is not zero-copy.

A spinlock implemented using C11 atomics protects producer insertion and exchange of
the two pool pointers. After a successful insertion, the producer sets an
atomic data-available flag with release ordering, then releases the lock.
The consumer drains its assigned pool without acquiring this lock. Only
when dequeue reports that this pool is empty does it inspect the flag.
If data are pending, the consumer acquires the shared lock, resets the flag,
exchanges the pool pointers, and releases the lock. The formerly empty
consumer pool becomes the new producer pool. No timer, minimum batch size,
or full-pool threshold triggers an exchange: it occurs when the consumer
is ready and producer data are available.

When both pools contain no pending work, the consumer remains in a polling
loop that checks whether producer data have become available. There is no
condition-variable or semaphore wake-up for this path. In contrast, the
capture loop blocks in \texttt{poll} on the netlink socket and shutdown pipe.
The polling design avoids a blocking wake-up in the consumer path, but does
not establish negligible idle CPU usage.

\subsubsection{Consumer Delay and Overflow}

While the consumer processes a batch, the producer continues filling its
current pool; it does not force an early swap or wait for the batch to
finish. It can nevertheless spin while the consumer holds the
handoff lock. If no free node remains in the producer pool, the default
policy removes that pool's oldest queued record and reuses the released
node for the new record. Free nodes in the consumer-owned pool are not
borrowed before the next ownership exchange. Consequently, the producer's
immediate buffering headroom is limited by its own $N$-node pool, not by
the combined free space of both pools. A slow callback or output operation
can therefore cause discards despite the separation of the two workers.

\subsubsection{Shutdown and Pending Records}

On receipt of an interrupt signal, the signal handler sets the shutdown flag and writes to the pipe,
waking the main thread and the netlink polling loop. The main thread
requests CSI disablement, joins the capture and consumer workers, unloads
the plugin, removes the monitor interface, and attempts to restore the
original station interface before closing the capture file and releasing
resources. The consumer tests the shutdown flag at its outer loop boundary;
it may finish a record already being processed, but does not explicitly
drain either pool before exiting. Queued records can therefore remain
unwritten at shutdown. Thread joining ensures worker termination, not
queue drainage, and closing the file does not process those remaining
records. Shutdown should accordingly be distinguished from a drain-and-flush
operation.

\subsection{Record Validation and Filtering}

Each record is validated before admission to an object pool. The capture path
verifies the netlink header, declared payload length, and maximum supported
payload size. Optional filters constrain the transmitter MAC address, channel
width, PHY format, and receive and transmit stream counts. An unrestricted mode
disables all the filters. After copying a structurally valid
payload, the capture handler assigns a microsecond-resolution timestamp
from the system's real-time (wall-clock) clock, before the remaining checks
and pool insertion. This is a userspace capture timestamp, not the time
at which the consumer writes the record.

The consumer invokes the active plugin callback synchronously on its local
CSI object and then, if base-file output was requested, emits the header and
payload using \texttt{writev}. Plugin processing and base-file output can
therefore coexist; selecting a plugin does not itself disable the base file.
The scatter/gather write avoids constructing another contiguous serialization
buffer. The implementation checks the returned byte count and logs a warning
at the configured verbosity on a short or failed write, but does not retry
the unwritten remainder. Bounded queue storage therefore does not imply
guaranteed record delivery to the output file.

\subsection{Plugin-Based Research Workflows}

The plugin interface lets experiments extend acquisition without modifying
the capture core. Plugins can inspect accepted CSI records, monitor conditions,
generate custom outputs, or coordinate external controls. Loaded at startup
through the versioned SDK, they support experiment-specific workflows.
Section~\ref{sec:plugin-system} describes this interface and demonstrates it
with infrared-controlled activity labeling.

\subsection{Inherited Issues and Implementation Corrections}

Review of the inherited FeitCSI acquisition path identified four avoidable
sources of per-record memory and I/O overhead \cite{feitcsi}.

First, FeitCSI dynamically allocated individual CSI objects without releasing
them after use, allowing memory consumption to grow. VesperCSI instead uses
preallocated, fixed-capacity pool slots and reuses them, avoiding per-record
heap allocation and bounding capture-buffer memory.

Second, the inherited processing path converted every raw CSI payload into
separate complex-value, magnitude, and phase arrays, although these arrays
served only the plotting path. The coefficient-by-coefficient transformations
were performed even when plotting was not needed. VesperCSI keeps these
plotting-specific calculations out of its acquisition core; a plugin or
downstream workflow can add them when needed.

Third, FeitCSI opened and closed the output file for every record and wrote
the header and payload separately. VesperCSI keeps the output file open
during a capture session and emits the header and payload together from
their existing buffers in a single vectored write. This reduces repeated
file-opening and write-call overhead without requiring an extra serialized
copy.

Finally, FeitCSI printed a message for every arriving record and provided no
way to silence this per-record output. The resulting console I/O added
work to the capture path. VesperCSI provides a command-line quiet option to
suppress routine per-record messages.

\section{Plugin System and Infrared-Controlled Activity Labeling}
\label{sec:plugin-system}

VesperCSI plugins are Linux shared-object libraries that extend the acquisition
program through a small, versioned interface. A plugin is compiled separately
against the VesperCSI plugin SDK and loaded by the program's dynamic loader at
startup, before acquisition begins. This is runtime loading rather than
compile-time inclusion: a compatible plugin can be built or replaced without
rebuilding or relinking the acquisition executable. The loader checks for the
required initialization, per-record callback, and destruction entry points.
Initialization provides the plugin with its SDK version, logging facilities,
and optional configuration path; destruction allows it to stop its workers
and release resources when acquisition ends.

The per-record callback is invoked synchronously by the consumer for each CSI
record admitted to the consumer path. It receives the current CSI object and
may inspect it, attach experiment metadata in plugin-managed output, or perform
other experiment-specific processing. Because the callback runs on the
consumer thread, it is part of the record-processing path rather than an
independent background task. If callback work takes longer than records arrive,
the capture producer continues filling its own preallocated pool while the
consumer is delayed. When that pool is exhausted, the default overflow policy
discards its oldest queued record to admit a new one. Thus, a slow plugin can
increase loss.

\begin{figure}[!t]
\centering
\includegraphics[width=\columnwidth]{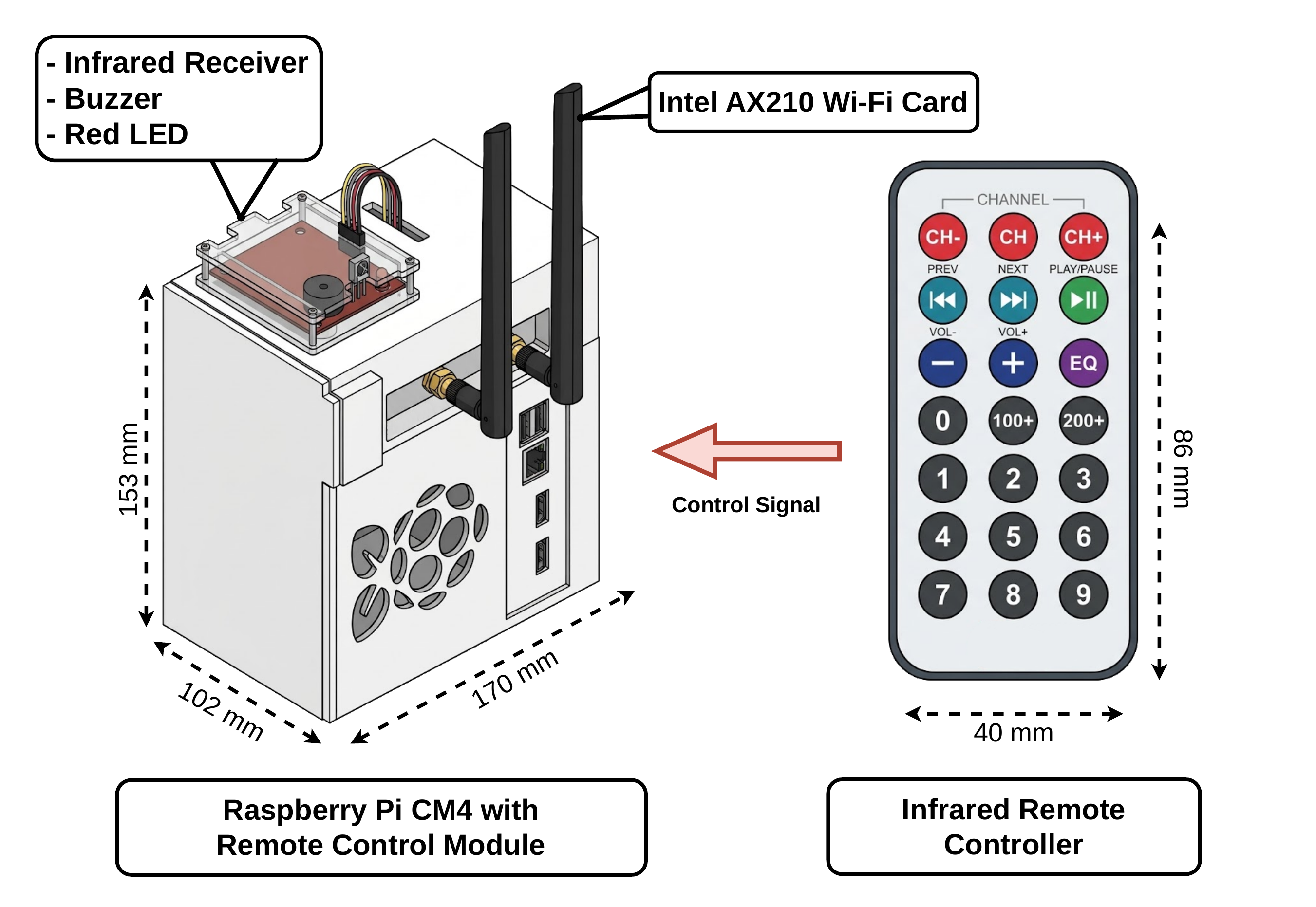}
\caption{Infrared-controlled activity-labeling setup. The Raspberry Pi CM4
case measures 153~mm in height (excluding the top module and antennas),
170~mm in width, and 102~mm in depth. The handheld remote controller measures
86~mm in height, 40~mm in width, 7~mm in depth; its depth extends perpendicular
to the image plane and is not visible in the two-dimensional figure. The
remote sends commands to an infrared receiver in a custom module attached to
the CM4 through GPIO. The module provides visual
and audible feedback during data collection.}
\label{fig:plugin-system}
\end{figure}

The most feature-rich plugin developed for VesperCSI to
date is its infrared-controlled activity-labeling plugin.
The infrared-controlled labeling plugin demonstrates how this interface can
combine CSI acquisition with a practical human-activity collection workflow.
As shown in Fig.~\ref{fig:plugin-system}, a custom module attached to the
Raspberry Pi CM4 provides an infrared receiver and GPIO-controlled indicators.
An operator uses a handheld remote to control capture and select activity
labels without interacting with the receiver through a separate computer
interface. Commands can start, stop, or discard a capture, and change the current
activity. The module's LED and buzzer
provide visual and audible feedback for capture and labeling state.

The plugin tracks the capture state, the activity being performed, and the
repetition number. While CSI is being recorded, it labels each measurement with
the current activity and repetition, then stores that information alongside
the CSI data in the same output stream. Users can configure activity durations,
pauses, and the number of repetitions,
supporting both operator-led and scheduled collection. Keeping these labels and
collection boundaries with the recorded data reduces the need to reconstruct
them from separate notes later.

The plugin interface is not limited to activity labeling. Any extension that
conforms to the SDK lifecycle and callback contract can add experiment-specific
monitoring, storage, or control behavior. It therefore allows VesperCSI to
support new research workflows without requiring changes to its core capture
path, within the execution and buffering limits of the plugin model.

\section{Positioning for ARM Edge Acquisition}
\label{sec:edge-positioning}

\tool{} targets the intersection of Intel AX200/AX210 CSI acquisition,
aarch64 edge deployment, and Linux 6.x compatibility. Its reported Linux 6.8
and 6.12 deployments demonstrate this combination, subject to a compatible
CSI-enabled \texttt{backport-iwlwifi} build. These tested versions do not imply
support for every Linux 6.x or subsequent kernel. Within the reviewed
landscape, the relevant AX200/AX210 host-acquisition platforms are PicoScenes,
IAX, FeitCSI, and \tool{}; matching the radio alone does not establish that
all four can serve as baselines on the same ARM testbed.

The deployment requirements narrow this set. PicoScenes' installation manual
requires amd64 and explicitly states that ARM processors are unsupported
\cite{picoscenesinstall}. Its exclusion from an ARM benchmark is therefore a
host-architecture constraint, not a lack of modern-kernel support or Intel AX
capability. IAX specifies a Linux 5.15.x API baseline for Intel AX extraction
and explicitly identifies Ubuntu's Linux 6.8 configuration as incompatible
with that baseline \cite{iaxrepo}. Using IAX would consequently require a
different kernel or additional porting, rather than a direct substitution
within the Linux 6.x testbed considered here.

Other acquisition platforms remain relevant to the broader sensing landscape
but address different hardware configurations. ESP-CSI targets ESP32-family
radios, Nexmon CSI targets Broadcom/Cypress devices, and AX-CSI and ZTECSITool
target access-point hardware \cite{nexmonrepo,axcsiproject,espcsi,ztecsirepo}.
The Linux 802.11n and Atheros tools use earlier-generation radios, while
openwifi uses an FPGA/SDR platform \cite{atheros,intelcsitool,openwifi}.
Benchmarking these systems would change the radio or execution model as well
as the acquisition software. Such comparisons can inform platform selection,
but cannot isolate userspace acquisition behavior on an Intel AX-based ARM host.

FeitCSI is therefore the closest directly applicable baseline among the
reviewed tools: it supports Intel AX200/AX210 acquisition and advertises aarch64
availability \cite{feitcsi}, and both programs were evaluated on CM4 hosts
running Linux 6.12, with the device-specific driver stacks described in
Section~\ref{sec:performance}. \tool{} emphasizes bounded
preallocated buffering, separation of reception from consumer work, and a
loadable interface for research workflows; FeitCSI offers CLI/GUI operation
and source-level customization without a documented equivalent plugin API
\cite{feitcsi}. These design differences motivate comparison under constrained
CPU resources, but do not alone establish a performance advantage. Section~\ref{sec:performance}
evaluates that question within the shared testbed rather than asserting a
general ranking across CSI platforms.

\section{Benchmarking Burst-Traffic and Long-Duration CSI Acquisition}
\label{sec:performance}

The evaluation consists of three experiments. The first compares the
burst-traffic acquisition performance of \tool{} and FeitCSI. The second
examines FeitCSI memory growth and swapping at 600~MHz and records a
higher-ceiling acquisition-process crash under the same extended
workload. The third evaluates \tool{} over repeated hour-scale sessions to
characterize sustained sequence retention and resource use.

\subsection{Receiver Platforms}
\label{sec:receiver-platforms}

The evaluation compared \tool{} and FeitCSI on two identical Raspberry Pi
Compute Module 4 hosts. Each contained a Broadcom BCM2711 quad-core
Cortex-A72 (ARMv8, 64-bit) SoC rated at 1.5 GHz, 2 GB LPDDR4-3200 SDRAM,
and 8 GB eMMC storage. Both receivers used Intel AX210 Wi-Fi cards for CSI
acquisition and Debian 13 (trixie). The receivers operated simultaneously at
equal distances from the transmitter. Device~A used Intel's public
\texttt{iwlwifi} stack, whereas device~B used the driver distributed with
FeitCSI. Table~\ref{tab:receiver-software} records the exact driver, kernel,
compiler, and userspace versions used in the evaluation.

\begin{table*}[!t]
\centering
\caption{Receiver software configurations used in the evaluation.}
\label{tab:receiver-software}
\scriptsize
\setlength{\tabcolsep}{3pt}
\renewcommand{\arraystretch}{1.0}
\begin{tabular*}{\textwidth}{@{\extracolsep{\fill}}>{\raggedright\arraybackslash}p{3cm}
                    >{\raggedright\arraybackslash}p{6cm}
                    >{\raggedright\arraybackslash}p{6cm}@{}}
\toprule
\textbf{Component} & \textbf{Device A} & \textbf{Device B} \\
\midrule
Driver & Intel \path{iwlwifi-stack-public:master:14952:c9d28532} &
FeitCSI \texttt{iwlwifi} 2.0.0 \\
Linux kernel & \path{6.12.47+rpt+rpi-v8} &
\path{6.12.75+rpt+rpi-v8} \\
Distribution & \texttt{Debian 13} & \texttt{Debian 13} \\
GCC & \texttt{14.2.0} & \texttt{14.2.0} \\
FeitCSI userspace & \texttt{2.0.0} & \texttt{2.0.0} \\
\tool{} userspace & \path{1.0.0-alpha.83} &
\path{1.0.0-alpha.83} \\
\bottomrule
\end{tabular*}
\end{table*}

Both receivers therefore used the same Linux major and minor kernel version,
6.12, while retaining the different patch and Raspberry Pi build revisions
shown in Table~\ref{tab:receiver-software}. Experiments~1 and~2 wrote their
capture files to the receivers' onboard eMMC storage. During Experiment~2,
the configured swap area was RAM-backed rather than mapped to the eMMC or
another secondary-storage device. Consequently, SwapPSS in that experiment
describes logical process pages placed in the RAM-backed swap mechanism and
does not represent secondary-storage traffic.

All \tool{} evaluation runs used the default worker-scheduling and overflow
behavior. CPU affinity, and real-time scheduling were not enabled. If the producer-owned pool
became full, the default policy discarded its oldest queued record and reused
that node for the incoming record.

Table~\ref{tab:pool-configuration} records the compile-time pool-sizing
parameters used for the \tool{} benchmark build. The derived quantities
follow the storage model in Subsection~\ref{sec:pool-layout}. They describe
the reserved CSI header and payload capacity rather than measured resident PSS and exclude
node metadata, pointers, alignment padding, and other process allocations.
The resulting estimate is 41,760,000~B (39.83~MiB) per pool and
83,520,000~B (79.65~MiB) across both pools.

\begin{table}[!t]
\centering
\caption{\tool{} pool configuration used in the evaluation.}
\label{tab:pool-configuration}
\scriptsize
\setlength{\tabcolsep}{3pt}
\renewcommand{\arraystretch}{1.0}
\begin{tabular}{@{}>{$}l<{$} l l@{}}
\toprule
\textbf{Quantity} & \textbf{Definition} & \textbf{Value} \\
\midrule
R_{\max} & Nominal sizing rate & 1000 records/s \\
T_{\max} & Nominal buffering window & 10 s \\
A_{\mathrm{tx}} & Maximum transmit paths & 2 \\
A_{\mathrm{rx}} & Maximum receive paths & 2 \\
S_{\max} & Maximum subcarriers & 244 \\
H & Header storage & 272 B \\
N & $R_{\max}T_{\max}$ & 10,000 \\
D & $4S_{\max}A_{\mathrm{tx}}A_{\mathrm{rx}}$ & 3,904 B \\
C & $H+D$ & 4,176 B \\
P & $NC$ & 39.83 MiB \\
P_{\mathrm{total}} & $2P$ & 79.65 MiB \\
\bottomrule
\end{tabular}
\end{table}

\subsection{Traffic Generation \& Radio Characteristics}

Traffic was generated from an amd64 PC running Debian 13 using a
dual-antenna USB Wi-Fi adapter based on the MediaTek MT7612U chipset. The
adapter was configured in monitor mode, and a custom C program injected IEEE
802.11 probe-response management frames in VHT format. The injector requested
an 80-MHz bandwidth, MCS~0, two spatial streams, a long guard interval, and no
link-layer acknowledgment through its 22-byte radiotap header. MCS~0 was used
as a robust modulation and coding configuration to reduce PHY decoding or
frame-check-sequence failures that could otherwise appear as acquisition
losses. With the benchmark SSID \texttt{CSI-VHT}, each constructed probe
response comprised a 131-byte IEEE 802.11 frame before any hardware-added FCS,
or 153 bytes in the local injection buffer including the radiotap header. The
program controlled inter-frame timing according to a user-specified target
traffic rate, providing fine-grained and repeatable scheduling. The primary
channel was channel~36 at 5180~MHz in the 5-GHz band.

Probe responses were selected because preliminary tests showed that not all injected frames, such as clear-to-send (CTS), produce CSI reports at the receivers, whereas
probe responses did. The receiver/destination address was the dummy address
\texttt{03:00:00:00:00:01}, which was not assigned to either receiver. The two
least-significant bits of its first octet identify it as a locally administered
multicast address. A probe response does not solicit a client response, and
the multicast destination and explicit no-acknowledgment request prevented
acknowledgment-based retransmissions and did not intentionally elicit traffic
from surrounding devices. This also preserved the uniqueness required by the
benchmark. The source MAC address strategy is described in the following
subsection.

A \texttt{iw dev} command snapshot for the MT7612U transmitter reported a
transmit-power setting of 18.00~dBm. The benchmark software did not explicitly
set transmit power, so this was the value reported for the adapter under its
driver and regulatory configuration. It is a configured, driver-reported
setting, not an independent measurement of radiated power. Both
Raspberry Pi receivers were positioned approximately 2~m from the transmitter
and placed immediately side by side, with no intentional gap between the
receiver units.

\subsection{Sequence Numbers Inside MAC Addresses}

CSI data delivered to the userspace acquisition program do not contain an
injector-controlled field, other than a MAC address, that can be used for
sequence tracking. We therefore divide the injected source MAC address into a
fixed magic prefix and a configurable sequence-number suffix. The magic bytes
distinguish the benchmark traffic from unrelated traffic in the surrounding
environment, while the suffix is varied for each injected frame. The
boilerplate source address is \texttt{AA:BB:CC:DD:EE:FF}, and the number of
low-order bytes reserved for sequence numbering can be selected according to
the experiment. For example, with a two-byte sequence field, the address takes
the form \texttt{AA:BB:CC:DD:XX:XX}, where \texttt{XX:XX} represents an
unsigned 16-bit sequence number. Comparing the received sequence identifiers
with the expected injected sequence provides deterministic identification of
numbered packets that were dropped before being retained by the acquisition
program.

\subsection{CPU Constraints}

Before each experimental condition, both Raspberry Pi receivers were configured
with the required CPU-frequency ceiling through the Linux sysfs interface
\path{/sys/devices/system/cpu/cpu*/cpufreq/scaling_max_freq}, where the
wildcard selects cores 0--3. The ceilings used throughout the experiments were
600, 800, 1000, 1200, and 1500~MHz. The same ceiling was applied to all four
cores simultaneously, imposing a system-wide frequency constraint rather than
limiting individual cores independently.

\subsection{General Benchmark Methodology}
\label{sec:benchmark-methodology}

\begin{figure}[!t]
\centering
\includegraphics[width=\columnwidth]{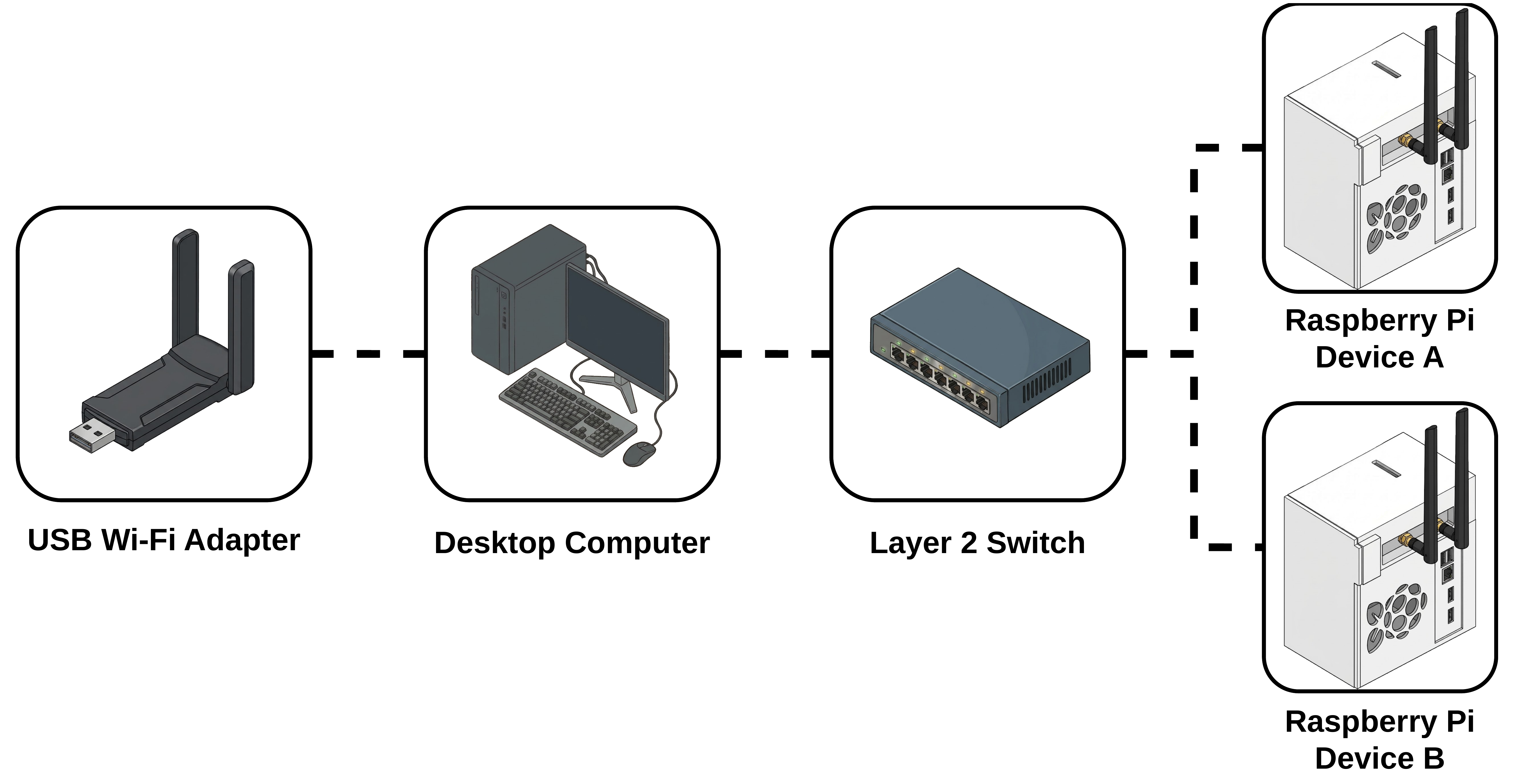}
\caption{Benchmark testbed topology. Two Raspberry Pi CM4 receivers connect
over Ethernet through a Layer 2 switch to the desktop computer, which controls
the trials. The USB Wi-Fi adapter attached to the desktop injects the wireless
benchmark traffic.}
\label{fig:testbed}
\end{figure}

The two Raspberry Pi receivers are connected through an Ethernet switch to
the main computer, and each receiver has a fixed IP address. The USB Wi-Fi
adapter used for frame injection is connected directly to the main computer.
Figure~\ref{fig:testbed} shows this arrangement, which separates the wired
control network from the wireless benchmark traffic. A Bash control script
running on the main computer serves as the benchmark controller. It issues
commands to the receivers over SSH and coordinates acquisition, resource
monitoring, sequence verification, and collection of results.

Before each campaign, the controller records the selected configuration and
checks remote access, required utilities, CPU-frequency controls, and
resource-sampling interfaces. It executes the configured CPU-ceiling and
traffic-rate combinations sequentially, repeating each condition as specified.
Either one or both receivers can be enabled, supporting single-tool experiments
and paired acquisition within the same workflow. When both receivers are
active, they observe the same injected traffic. Frame counts, software
assignments, repetition counts, and timing parameters are specified separately
for each experiment.

Each trial begins with system CPU and temperature sampling to establish a
pre-acquisition baseline. The controller launches a new acquisition process
on each enabled receiver and initiates process CPU and memory sampling.
Acquisition processes are launched sequentially over SSH; injection begins
after a startup interval and a process-liveness check. CSI records are written
locally on the receivers. During acquisition, background samplers record
process CPU use, resident and swapped process memory, per-core system CPU
activity, and SoC temperature. Elapsed times are measured relative to each
receiver's trial baseline.

Following completion of frame injection, the controller waits for the
configured post-injection interval, intended to give any frames queued in the
transmitter's kernel transmit path time to drain. This fixed delay does not
verify receiver-side delivery or confirm that CSI records have reached the
capture files. The controller then checks acquisition-process liveness and
invokes a sequence-verification program on the capture file of each receiver
whose acquisition process remains active. The program compares the sequence
identifiers encoded in source MAC addresses with the expected sequence and
reports unique receptions, missing identifiers, duplicates, and out-of-range
identifiers. Following verification, acquisition processes are stopped, with stronger
termination signals used when necessary. System CPU and temperature sampling
extends briefly beyond process termination to provide post-acquisition
context. Configuration snapshots, injection and acquisition logs, verification
output, and resource traces are collected on the main computer and organized
by condition and repetition. A configurable cooldown interval can be applied
between trials. Acquisition-process failures are recorded separately from
packet loss. A receiver whose acquisition process fails is excluded from
subsequent trials, while acquisition on any remaining active receiver continues.

\subsection{Experiment 1: Burst-Traffic Comparison}
\label{sec:experiment1}

Experiment~1 compares the retention of numbered CSI records and the associated
host-resource demands under short, finite traffic workloads. Two campaigns
exchange the acquisition programs between the receivers: assignment~I places
\tool{} on device~A and FeitCSI on device~B, whereas assignment~II places
FeitCSI on device~A and \tool{} on device~B. Within each campaign, both
receivers observe the same injected burst. Each of the five all-core CPU
ceilings, 600, 800, 1000, 1200, and 1500~MHz, is combined with seven target
rates, 100, 200, 500, 1000, 1500, 2000, and 2500~frames/s. Ten consecutive
repetitions of each condition inject 10,000 frames, with four source-MAC
bytes reserved for sequence numbering. The resulting dataset contains
350 paired trials per assignment, or 700 injected bursts and 1400 receiver
observations in total. All injector logs report 10,000 successful
\texttt{sendto} calls to the transmitter interface. This software-side count
does not verify over-the-air transmission or reception; receiver captures are
assessed separately by sequence verification. Both campaigns contain
complete verification results without recorded collector crashes. The
configured startup and post-injection drain intervals are 5 and 10~s,
respectively. The nominal injection duration consequently varies from
approximately 100~s at 100~frames/s to 4~s at 2500~frames/s.
The configured inter-frame intervals are 10,000, 5000, 2000, 1000, 667,
500, and 400~$\mu$s, respectively; rounding to integer microseconds
gives 1499.25~frames/s for the nominal 1500-frames/s condition.

The crossover retains the hardware and driver associated with each device;
it exchanges the userspace tools rather than independently varying the
drivers. Agreement between assignments therefore supports the robustness
of the observed ordering on these two device/driver combinations, but does
not isolate the contribution of each acquisition component. The campaigns
are sequential, and the repeated conditions are not randomized. Results
below are descriptive summaries of the retained trials, without an assumption
that consecutive repetitions provide independent realizations of the
wireless environment.

\subsubsection{Sequence Retention Across CPU and Traffic Conditions}

For trial~$j$, the verifier counts the distinct expected sequence identifiers
from 0 through 9,999 found in each receiver's capture file; denote this count
by $U_j$. The missing-record count is $D_j=10{,}000-U_j$, and the corresponding
loss percentage is $100D_j/10{,}000$. Thus, retention measures which expected
identifiers appear in the saved capture, not where an absent frame was lost.
No duplicate or out-of-range identifiers are reported in either campaign.
Table~\ref{tab:exp1-loss} gives the mean missing-record count
and its sample standard deviation across ten repetitions for every
CPU/rate combination. The standard deviations describe
variation between trials. Sequence accounting does not establish whether a
missing identifier was lost over the radio link, in firmware, in the driver,
or in userspace.

\input{tables/experiment1_packet_loss}

\begin{figure*}[!t]
\centering
\includegraphics[width=0.95\textwidth]{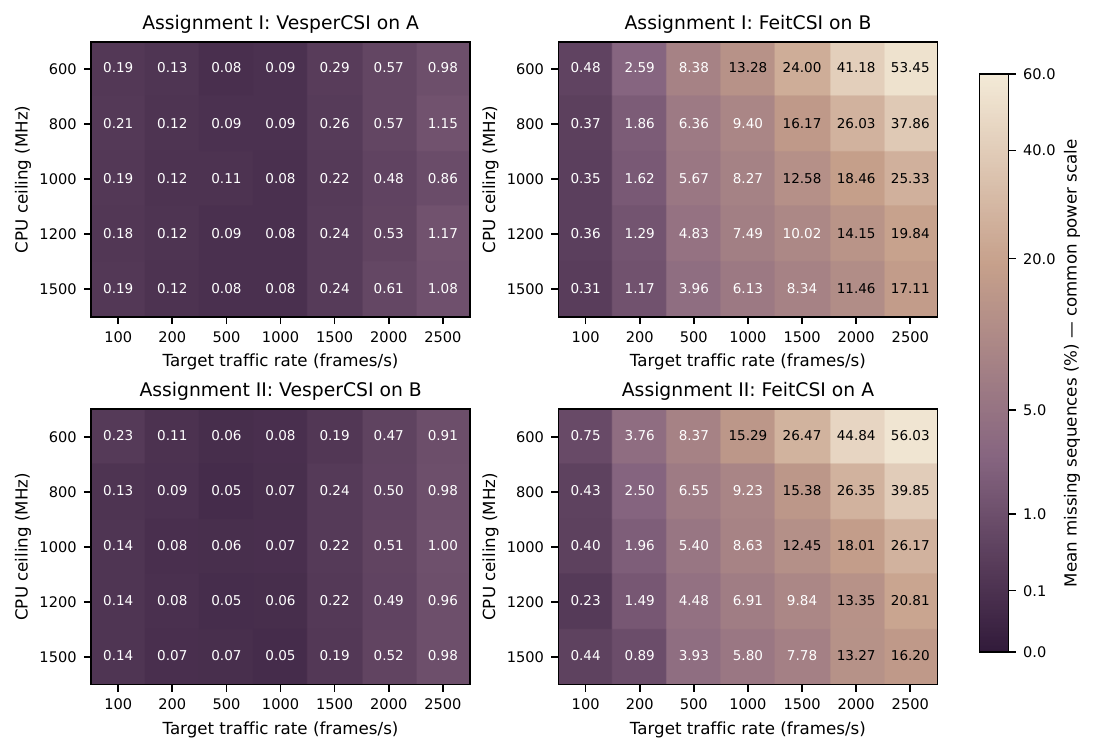}
\caption{Mean missing-sequence percentage over ten repetitions for every
Experiment~1 condition. Rows distinguish the two software assignments and
columns distinguish the tools. Cell annotations provide the percentages;
all panels share the same nonlinear color scale to retain visibility of
sub-percent losses. Labels A and B denote Device~A and Device~B, respectively;
the two Raspberry Pi CM4 receivers are shown in Fig.~\ref{fig:testbed}.}
\label{fig:exp1-loss}
\end{figure*}

Figure~\ref{fig:exp1-loss} shows that \tool{} has a lower mean missing-record
percentage in all 35 configurations under each assignment. Its condition
means range from 0.077 to 1.166\% in assignment~I and from 0.046 to
0.999\% in assignment~II. FeitCSI's corresponding ranges are
0.313--53.447\% and 0.231--56.028\%. At the most constrained high-rate
condition, 600~MHz and 2500~frames/s, \tool{} loses a mean of 98.3 records
(0.983\%) on device~A and 91.3 records (0.913\%) on device~B. FeitCSI
loses 5344.7 records (53.447\%) on device~B and 5602.8 records
(56.028\%) on device~A. Thus, the large separation at this condition
persists after exchanging the programs between devices.

Increasing the CPU ceiling reduces FeitCSI's mean loss at the highest
traffic rate: at 1500~MHz and 2500~frames/s, the percentages fall to
17.113\% and 16.199\% in assignments~I and~II, respectively. The same
conditions yield 1.080\% and 0.984\% for \tool{}. \tool{} also exhibits higher losses at the upper traffic
rates than at several lower-rate conditions, and its response to CPU
frequency is not uniformly monotonic. The results therefore demonstrate
improved retention within the tested configuration rather than lossless
capture or a universal performance bound. At 100~frames/s, the absolute
separation is much smaller, with both tools losing less than 0.8\% on
average in every condition.

\subsubsection{Process CPU Occupation and Memory Footprint}

Resource traces are summarized using each receiver's local elapsed time and
the controller's recorded active phase. This phase extends from collector
startup to the post phase after collector termination and includes startup,
injection, drain, verification, and termination overhead. The traces do not
provide separate injection boundaries, so these summaries are not restricted
to the interval during which frames are transmitted. CPU and memory sampling
are configured at 10~Hz, although sampler execution and scheduling add
overhead to the nominal sampling interval. For each trial, mean process CPU
occupation is computed from the change in process CPU ticks divided by the
elapsed span between its first and last active process samples. Values are
normalized so that 100\% denotes one fully occupied logical core. All four
cores remain available with the same frequency ceiling; the recorded launch
commands impose no single-core affinity. CPU normalization therefore does
not imply that the process is pinned to one core.

Memory measurements are proportional set size (PSS) and proportional swapped
memory (SwapPSS) from \path{/proc/PID/smaps_rollup}. For each trial, the
memory figure and the $M$ columns of
Tables~\ref{tab:exp1-resources} and~\ref{tab:exp1-resources-II} use
the sample mean of PSS + SwapPSS, with both values
summed at each sampling instant and converted from KiB to MiB. This combined
metric accounts for process-attributed resident and swapped pages, including
proportional shares of shared pages; it does not measure simultaneously
occupied physical RAM.
Tables~\ref{tab:exp1-resources} and~\ref{tab:exp1-resources-II} report
assignments~I and~II separately, with one row for each of the 35 CPU-ceiling
and target-rate combinations. Each entry gives mean $\pm$ sample standard
deviation of ten trial summaries for CPU occupation, mean combined memory,
or baseline-referenced temperature rise. The SD uses denominator $10-1$
and describes repetition variability within that configuration; trials from
different traffic rates are not pooled. The resource figures show the same
condition means and between-trial sample SDs. These SDs are not standard
errors, confidence intervals, or variation among samples within a trace.

\input{tables/experiment1_resources}

\begin{figure*}[!t]
\centering
\includegraphics[width=0.95\textwidth]{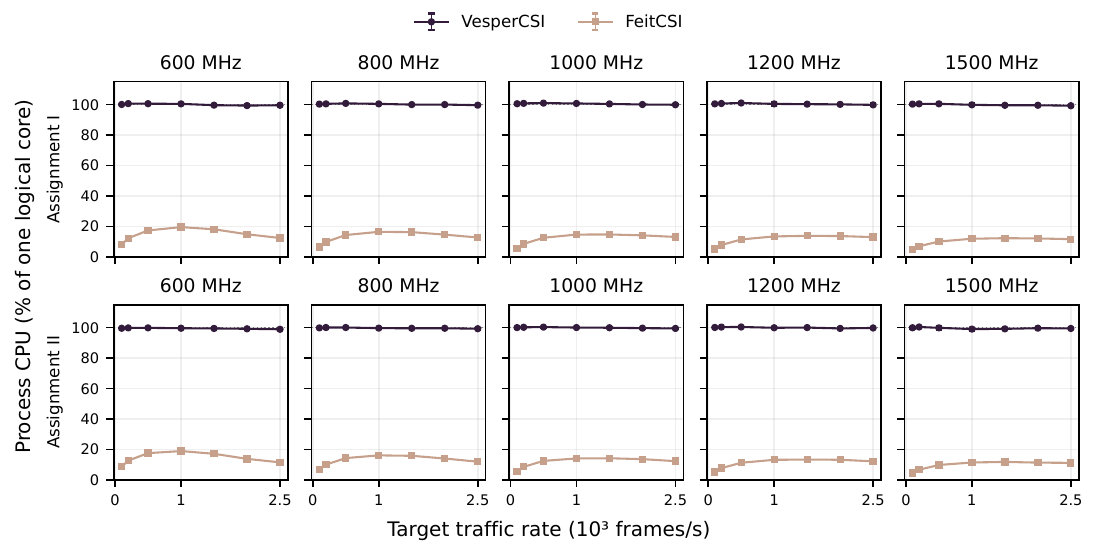}
\caption{Mean process CPU occupation during the observed active process
samples, normalized to one logical core. Columns identify the all-core CPU
ceilings and rows identify assignments~I and~II. Points show means across
ten trials; error bars show their sample standard deviations. The dotted
reference denotes one fully occupied logical core.}
\label{fig:exp1-cpu}
\end{figure*}

Figure~\ref{fig:exp1-cpu} shows that \tool{} occupies approximately one
logical core throughout the sweep: its condition means span
98.86--100.99\% across the two assignments. FeitCSI's condition means
span 4.51--19.53\%. At 600~MHz and 2500~frames/s,
Tables~\ref{tab:exp1-resources} and~\ref{tab:exp1-resources-II} give
\tool{} CPU occupation of $99.52\pm0.21$\% in assignment~I and
$98.86\pm0.29$\% in assignment~II, compared with
$12.39\pm0.38$\% and $11.51\pm0.26$\% for FeitCSI, respectively.
The higher CPU occupation of \tool{} accompanies
its improved retention but also represents a resource cost that remains
visible at the lower traffic rates. Conversely, the lower active-phase
occupation of FeitCSI does not imply a more efficient acquisition path:
the tools retain different numbers of records, and the measurement includes
periods without injected traffic. These averages cannot establish the
CPU service time per retained record or explain FeitCSI's losses. Small
deviations around 100\% can reflect multithreaded execution and the finite
resolution of CPU-tick and elapsed-time measurements.

\begin{figure*}[!t]
\centering
\includegraphics[width=\textwidth]{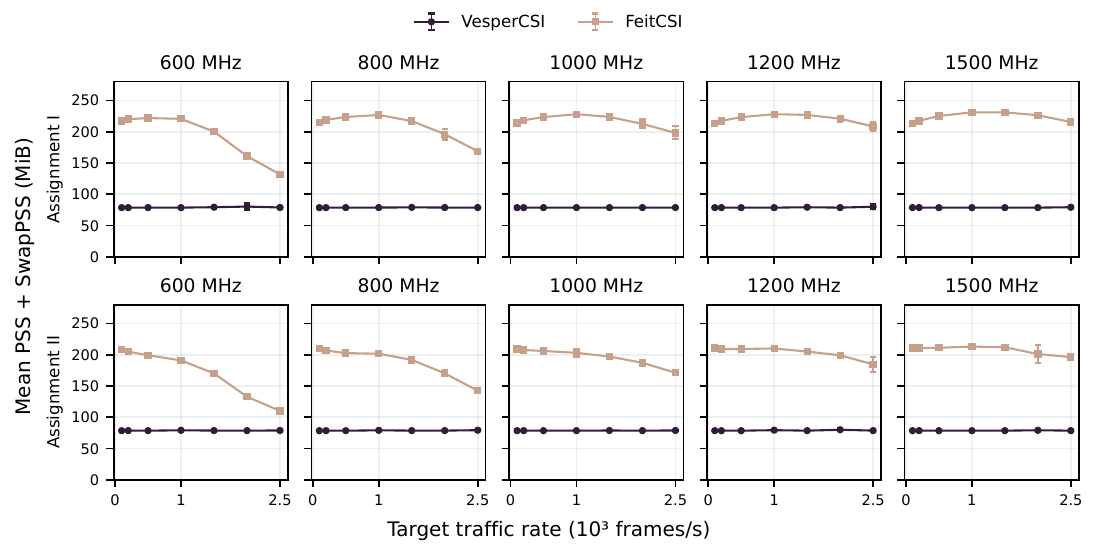}
\caption{Sampled mean combined process memory (PSS + SwapPSS) during each receiver's active phase,
in MiB, for every CPU ceiling and traffic rate. Points and error bars denote
the mean and sample standard deviation of ten trial means, calculated by
summing resident and swapped memory at each sampling instant. The metric
excludes other host memory and does not
represent physical RAM occupancy.}
\label{fig:exp1-memory}
\end{figure*}

Figure~\ref{fig:exp1-memory} shows a substantially smaller and less
condition-dependent combined process-memory footprint for \tool{}. Its condition means
of trial-mean PSS + SwapPSS range from 78.63 to 80.38~MiB, compared with
110.22--230.98~MiB for FeitCSI. No nonzero SwapPSS samples occur in
either tool's active-phase traces in Experiment~1, so the combined values
equal the resident PSS values numerically, although the figure and tables
include both components. These ranges summarize ten-trial condition means
rather than bounds on individual trials. The burst results are consistent with a more stable
combined process-memory footprint for \tool{}, but do not independently identify an
allocation defect in FeitCSI or establish long-duration memory behavior.

Tables~\ref{tab:exp1-resources} and~\ref{tab:exp1-resources-II}
quantify the memory separation within each configuration. At 600~MHz
and 2500~frames/s, mean combined memory is $78.90\pm0.04$~MiB
for \tool{} and $131.80\pm4.29$~MiB for FeitCSI in assignment~I.
After exchanging the tools in assignment~II, the corresponding values
are $78.93\pm0.08$ and $110.22\pm2.20$~MiB. At 1500~MHz and
the same traffic rate, FeitCSI's values rise to $215.55\pm4.93$~MiB
and $196.22\pm4.90$~MiB, whereas \tool{} remains at
$79.32\pm1.48$~MiB and $78.74\pm0.05$~MiB. The SDs describe
the spread of the ten trial means at each fixed CPU ceiling and traffic
rate. Thus, the smaller overall variation in \tool{}'s condition means
does not imply an identical footprint in every repetition.

\subsubsection{Baseline-Referenced Temperature Rise}

Each retained temperature trace reports the \texttt{cpu-thermal} SoC
sensor, rather than separate sensors for the four cores. For trial~$j$,
the reported rise is
$\Delta T_j=\overline{T}_{j,\mathrm{active}}-
\overline{T}_{j,\mathrm{baseline}}$, using the sample mean in each
recorded phase. Baseline subtraction reduces sensitivity to the initial
temperature, while retaining the effects of trial duration, cooling,
background system activity, and thermal history.

\begin{figure*}[!t]
\centering
\includegraphics[width=\textwidth]{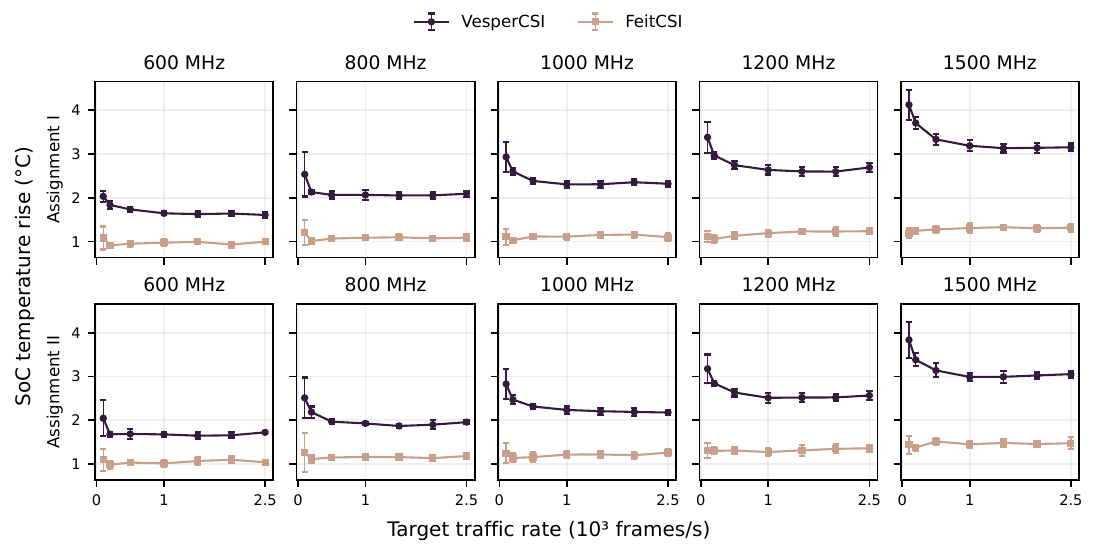}
\caption{Mean active-phase SoC temperature rise relative to each trial's
mean pre-acquisition baseline. Each point is the mean of ten baseline-referenced
trial rises; error bars show their sample standard deviation. The two rows
report the assignments separately. Active phases differ in duration across
traffic rates.}
\label{fig:exp1-temperature}
\end{figure*}

Figure~\ref{fig:exp1-temperature} shows a higher baseline-referenced
temperature rise for \tool{} than for FeitCSI in every condition mean.
Across both assignments, the rises span 1.61--4.13~$^\circ$C for
\tool{} and 0.91--1.51~$^\circ$C for FeitCSI.
Tables~\ref{tab:exp1-resources} and~\ref{tab:exp1-resources-II}
give the ten-trial mean and SD for each configuration. At 2500~frames/s,
increasing the CPU ceiling from 600 to 1500~MHz raises \tool{}'s mean
temperature rise from $1.61\pm0.06$ to $3.15\pm0.09$~$^\circ$C
in assignment~I and from $1.72\pm0.04$ to
$3.05\pm0.08$~$^\circ$C in assignment~II. FeitCSI's corresponding
values increase from $0.99\pm0.06$ to $1.32\pm0.09$~$^\circ$C
and from $1.04\pm0.05$ to $1.47\pm0.13$~$^\circ$C.
This thermal separation accompanies
the difference in CPU occupation, but does not constitute a direct
measurement of electrical power or energy efficiency. In particular,
lower-rate trials transmit the same 10,000 frames over a longer interval,
and the active phase includes control and drain time. Temperature differences
across traffic rates should therefore not be interpreted as steady-state
thermal responses to traffic intensity alone.

\subsection{Experiment 2: FeitCSI Memory Growth at 600 MHz}
\label{sec:experiment2}

Experiment~2 examines whether the resident-memory behavior observed in
short bursts remains suitable when the acquisition workload is extended.
This analysis focuses on FeitCSI on device~B with a 600-MHz frequency
ceiling applied to all four CPU cores. Ten consecutive trials each inject
100,000 numbered frames at a target rate of 1000~frames/s, corresponding
to a nominal 100-s injection interval. The startup and post-injection drain
intervals remain 5 and 10~s. A fresh acquisition process is launched for
each trial; the reported growth therefore occurs within individual
sessions rather than accumulating across successive processes. All ten
injector logs report 100,000 successful \texttt{sendto} calls, and all ten
600-MHz trials reach receiver-side sequence verification without a recorded
collector failure.

Table~\ref{tab:exp2-loss} gives the missing-record mean and sample standard
deviation across the ten completed 600-MHz, 1000-frames/s trials. FeitCSI
misses $45483.6\pm458.0$ of the 100,000 expected identifiers per trial,
equivalent to $45.484\pm0.458$\% loss. These SDs describe variation
between trials, not uncertainty in the mean or the location of losses
within the acquisition path.

\begin{figure*}[!t]
\centering
\includegraphics[width=\textwidth]{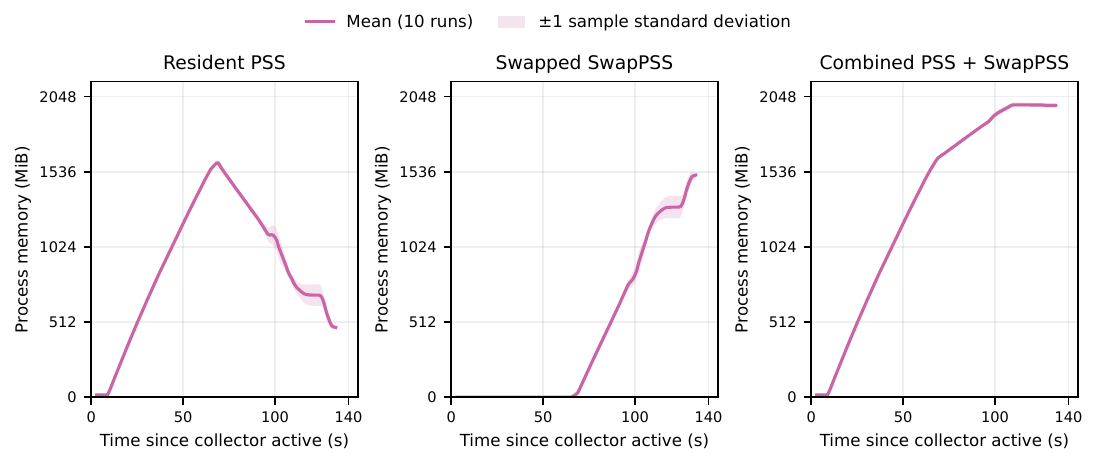}
\caption{Mean FeitCSI memory trajectories across the ten Experiment~2 trials at the
600-MHz all-core ceiling and 1000-frames/s target. The panels distinguish
resident PSS, swapped SwapPSS, and their sum, using a common memory scale.
Time is relative to each trial's recorded collector-active phase.
Solid curves show the mean across runs; shaded bands show $\pm1$ sample
standard deviation across runs at each time point. Traces are linearly interpolated onto a common 1-s grid from
3 to 133~s, restricted to the interval observed in all ten runs.
The trajectories include acquisition and subsequent control/drain overhead;
injection boundaries are not separately marked in the resource logs.}
\label{fig:exp2-memory-traces}
\end{figure*}

\input{tables/experiment2_packet_loss}

Despite retaining only approximately 54.5\% of the expected identifiers
on average, FeitCSI's combined process-memory demand approaches the installed-memory
capacity. The ten 600-MHz trials establish reproducible memory growth and
swapping. Motivated by the improved short-burst retention observed at higher
CPU ceilings in Experiment~1, we attempted the same 100,000-frame stress
workload at 1500~MHz. The acquisition process crashed
before usable resource or kernel traces were preserved. Consequently, this run
does not establish the 1500-MHz memory trajectory or the cause of failure. Its
occurrence alongside the repeated near-capacity behavior at 600~MHz is
consistent with memory pressure, but does not prove that memory pressure caused
the crash.

The memory sampler records proportional resident memory (PSS) and
proportional swapped memory (SwapPSS) from
\path{/proc/PID/smaps_rollup}, at a configured sampling rate of 10~Hz.
The figures use each receiver trace's local elapsed time, referenced to
the recorded start of the active phase, and convert KiB to MiB. As in
Experiment~1, this phase includes startup, injection, drain, verification,
and termination overhead. PSS describes resident pages attributed to the
process; SwapPSS describes attributed pages placed in the RAM-backed swap
area described in Subsection~\ref{sec:receiver-platforms}. Their sum
is used here as a combined process-memory demand. It is not a measurement
of simultaneously occupied physical RAM and excludes other processes
and kernel memory.

Figure~\ref{fig:exp2-memory-traces} shows substantial memory growth in
the mean trajectory, with the shaded bands describing variation between
sessions. At each plotted time, the mean and sample standard deviation
use ten equally weighted interpolated observations, with the standard
deviation calculated using nine degrees of freedom. For the combined
panel, PSS and SwapPSS are summed within each run before computing these
statistics; the two component standard deviations are not added.
The following ranges summarize the original individual traces rather
than extrema of the plotted mean. Initial sampled PSS is
14.51--15.31~MiB, whereas the maximum sampled resident PSS reaches
1572.13--1634.35~MiB across the ten trials. The first nonzero SwapPSS
sample occurs 65.10--68.60~s after the recorded active-phase start.
Thereafter, resident PSS decreases while SwapPSS increases, with the
maximum swapped footprint reaching 1488.45--1536.57~MiB. The reduction
in resident memory therefore accompanies substantial displacement to swap;
it does not indicate that the combined memory demand has returned to its
initial level. By the last available samples, approximately
133.70--137.79~s after the active-phase start, that demand remains close
to 2~GiB. The traces do not identify the allocation mechanism responsible
for this growth, but they directly establish a large process footprint
and sustained swapping during the tested workload.

\begin{figure}[!t]
\centering
\includegraphics[width=\columnwidth]{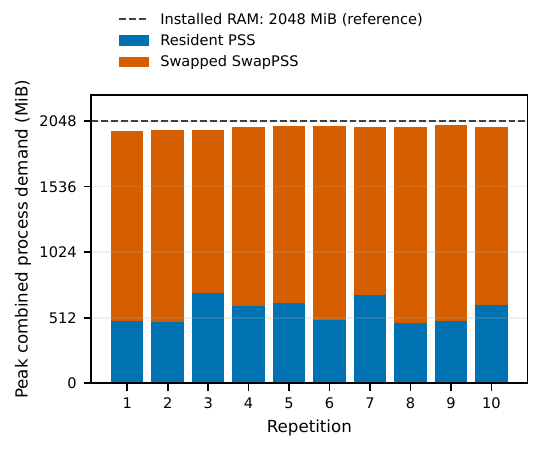}
\caption{Peak sampled combined FeitCSI memory demand in each 600-MHz
trial. Each stack uses PSS and SwapPSS from the same sample at which their
sum is maximal, rather than adding their separately observed maxima.
The dashed line is a nominal 2048-MiB installed-memory reference for the
2-GB platform; the stacked quantity includes swapped pages and is not
physical RAM occupancy.}
\label{fig:exp2-memory-peaks}
\end{figure}

Figure~\ref{fig:exp2-memory-peaks} quantifies the capacity pressure across
repetitions. Peak combined demands range from 1972.73 to 2019.24~MiB. Their
numerical ratio to the nominal 2048-MiB installed-memory reference is
96.32--98.60\%, but this is a scale comparison, not the fraction of physical
RAM occupied: the combined metric includes swapped pages. Separately, the
resident peaks alone amount to approximately 77--80\% of that reference,
before accounting for the operating system and other host workloads. The
substantial swapped component demonstrates
that completion of these finite trials depends on memory behavior beyond
a small, stable resident acquisition footprint. Process-memory samples do
not establish that FeitCSI alone occupies every byte of the Pi's RAM, but
the near-capacity combined demand and repeated swapping document severe
memory pressure on this platform.

For the tested FeitCSI build and acquisition configuration, these results
support restricting operation to short, bounded sessions rather than
relying on unattended long-duration capture on a 2-GB Raspberry Pi.
Each nominal 100-s traffic workload already produces a footprint of the
same order as the installed RAM, leaving little margin in the memory
budget for a larger workload or additional applications. Extended
deployment would require addressing the observed growth or applying
explicit session and resource limits. Nevertheless, the ten 600-MHz
trials complete, so this experiment does not establish a universal maximum
run duration or prove that every longer session must fail. It demonstrates
that stable memory use, a prerequisite for dependable long-duration edge
acquisition, is not achieved during the measured sessions.

\FloatBarrier

\input{tables/experiment3_retention}

\begin{figure*}[!t]
\centering
\includegraphics[width=0.95\textwidth]{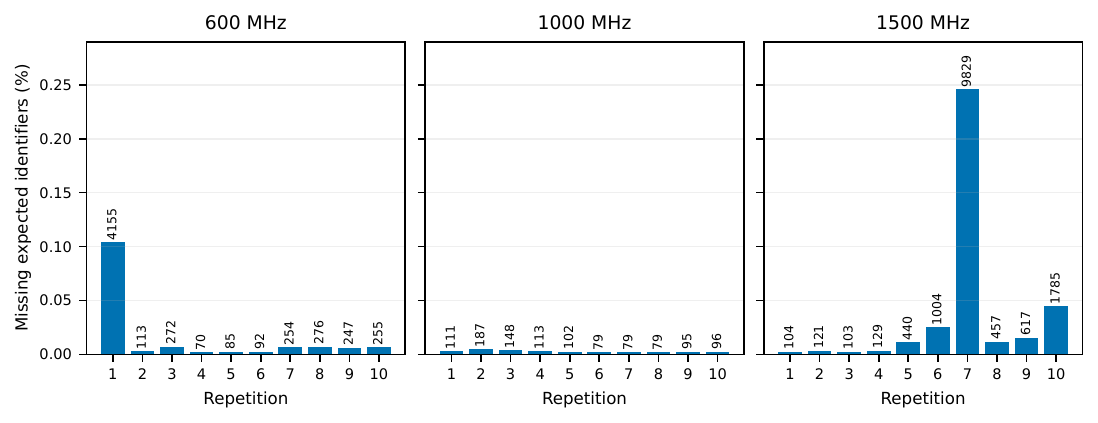}
\caption{Missing-sequence percentages for all thirty Experiment~3 trials,
grouped by the all-core CPU ceiling. Each trial expects four million
identifiers. Bar annotations give the corresponding missing-record counts,
including the larger-loss trials that affect the configuration averages.}
\label{fig:exp3-loss}
\end{figure*}

\subsection{Experiment 3: Sustained VesperCSI Acquisition}
\label{sec:experiment3}

Experiment~3 evaluates the ability of \tool{} to retain CSI records over
repeated hour-scale sessions while maintaining a small resident footprint.
Only device~A, running \tool{}, is enabled. The 600-, 1000-, and 1500-MHz
all-core CPU ceilings are each tested in ten consecutive trials, with four
million numbered frames injected per trial at a target rate of 1000~frames/s.
The injector encodes each frame's sequence number in the final four bytes of
its source MAC address. Each trial uses a fresh acquisition process, a 5-s
startup interval, and a 10-s post-injection drain interval.
To accommodate the storage requirements of the extended runs,
capture files are written to an 8-TB Western Digital My Book external drive
connected to the receiver via USB. A configured 900-s (15-min) cooldown
separates successive trials. All thirty injector logs report four million
successful \texttt{sendto} calls, every trial reaches receiver-side sequence
verification, and no collector crashes are recorded.

The nominal injection duration is 4000~s, or approximately 66.7~min per
trial. In the receiver captures, the earliest and latest timestamps attached
to matching expected identifiers are separated by 4000.023--4000.034~s.
Thus, matching records bracket the intended hour-scale workload; this
first-to-last span does not imply that every intervening identifier was
retained. Together, the trials cover
120 million expected identifiers and approximately 33.3~h of injected
traffic. This total represents thirty separate acquisition sessions with
cooldown, rather than one uninterrupted 33-h run. The recorded active
phases are longer, averaging 75.98, 75.27, and 74.86~min at 600, 1000,
and 1500~MHz, respectively, because they also include startup, drain,
verification, and termination overhead; these durations exclude the separate
cooldowns. Resource summaries below use the active phases and each receiver
trace's local elapsed time.

\subsubsection{Retention Throughout the Extended Workloads}

\input{tables/experiment3_resources}

\begin{figure*}[!t]
\centering
\includegraphics[width=0.9\textwidth]{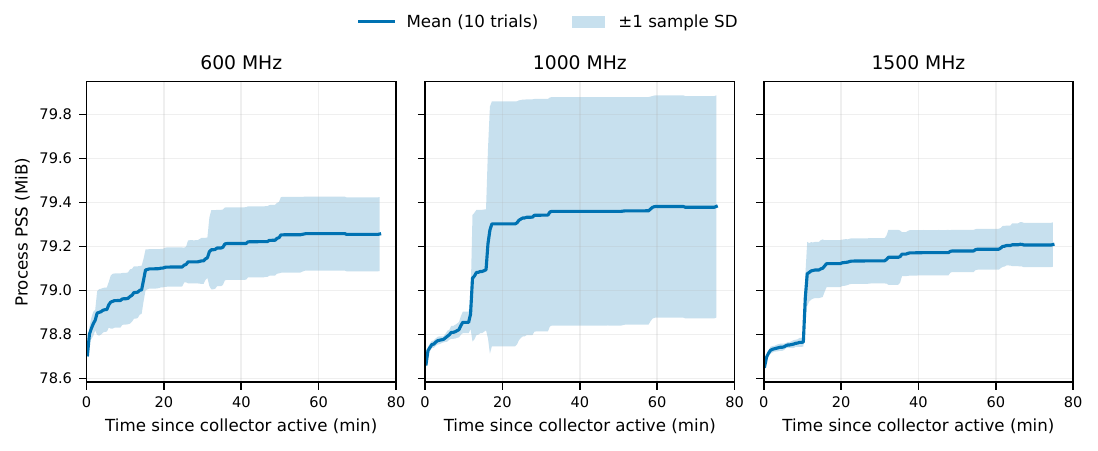}
\caption{Mean process PSS during the extended \tool{} trials, with
separate panels for the three CPU ceilings. Solid curves show the mean of
ten trial sample means in each 30-s bin; shaded bands show $\pm1$ sample
standard deviation across those ten values. Bins are shown only while all
ten trials contribute. The magnified vertical scale makes small footprint
changes visible.
Table~\ref{tab:exp3-resources} uses original samples for maxima.
No nonzero SwapPSS samples occur in any trial.}
\label{fig:exp3-memory}
\end{figure*}

Table~\ref{tab:exp3-retention} reports sequence accounting for every
configuration. Each row pools the ten equal-sized trials (40 million
expected identifiers per CPU ceiling); retention is 99.985453\% at 600~MHz,
99.997278\% at 1000~MHz, and 99.963528\% at 1500~MHz. Across all
thirty trials, 119,978,503 of the 120,000,000 expected identifiers are
retained, giving 99.982086\% aggregate retention and 21,497 missing
identifiers. No duplicate or out-of-range identifiers are recorded.
These counts establish high end-to-end retention of the expected
sequence; they do not locate individual losses within the radio,
firmware, driver, or userspace acquisition path.

Figure~\ref{fig:exp3-loss} shows that the high aggregate retention is
accompanied by variation between sessions. The first 600-MHz trial loses
4155 identifiers (0.103875\%), whereas the remaining nine lose
70--276 each. At 1000~MHz, all ten trials lose 79--187 identifiers,
with a worst-trial loss of 0.004675\%. At 1500~MHz, the seventh trial
loses 9829 identifiers (0.245725\%), and several later trials also
exceed the losses of the earlier repetitions. Even this largest-loss
trial retains 99.754275\% of its expected sequence. The observations
support sustained high retention across the tested hour-scale sessions,
including operation at 600~MHz, while also showing that retention is
neither perfect nor monotonically improved by a higher frequency ceiling.
The ordered repetitions do not isolate environmental drift or identify
the cause of the larger losses, so the aggregate result should be read
together with the individual trials.

\subsubsection{Resident Memory and CPU Occupation}

Table~\ref{tab:exp3-resources} summarizes the active-phase process and
temperature measurements for each frequency configuration. Mean CPU
occupation is calculated from accumulated process CPU ticks divided by
the observed elapsed sampling span and normalized to one logical core.
Memory summaries use the sampled PSS from
\path{/proc/PID/smaps_rollup}, converted to MiB; endpoint change is
the last active PSS sample minus the first. Reported $\pm$ values are
sample standard deviations across ten trial summaries, whereas sampled
maxima retain the largest original observation across the configuration.

\begin{figure*}[!t]
\centering
\includegraphics[width=\textwidth]{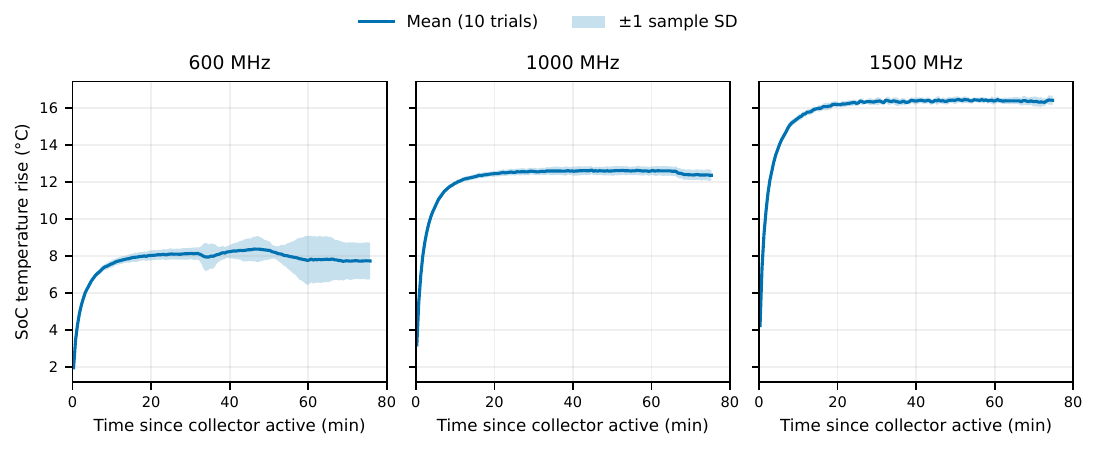}
\caption{Mean baseline-referenced SoC temperature rise across ten trials
of each Experiment~3 configuration. For each trial, 30-s active-temperature
sample means are referenced to that trial's mean baseline. Solid curves
show the mean of these ten trial values in each bin; shaded bands show
$\pm1$ sample standard deviation across trials. Bins are shown only while
all ten trials contribute.}
\label{fig:exp3-temperature}
\end{figure*}

\begin{figure*}[!t]
\centering
\includegraphics[width=\textwidth]{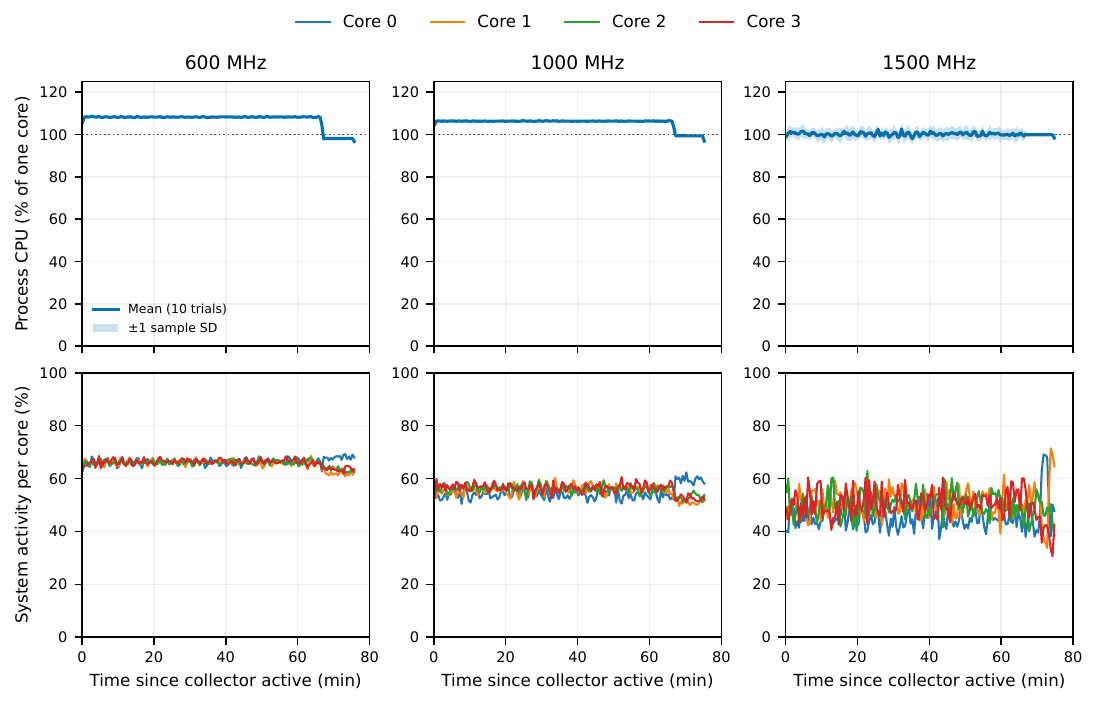}
\caption{Active-phase CPU behavior during Experiment~3. Upper panels
show the mean process CPU occupation across ten trials, normalized to one
logical core, with shaded $\pm1$ sample standard deviation bands. Each trial
contributes a duration-weighted mean in each 30-s bin; dotted lines mark
100\%. Lower panels show
per-core total system activity from \protect\path{/proc/stat}, averaged across
the ten trial means in each 30-s bin. Bins in both rows are shown only
while all ten trials contribute. System activity includes the acquisition process, samplers, and other host workloads.}
\label{fig:exp3-cpu}
\end{figure*}

Figure~\ref{fig:exp3-memory} shows a mean resident footprint that remains
close to 79~MiB throughout the sessions, with small changes
rather than growth on the scale of the acquired record count. The means
of trial-mean PSS are 79.15, 79.25, and 79.10~MiB at 600, 1000,
and 1500~MHz, respectively. Their mean first-to-last increases are
0.80, 0.93, and 0.76~MiB, and the largest original PSS sample in the
entire experiment is 80.79~MiB. Thus, the sampled process footprint
remains below 4\% of a nominal 2048-MiB installed-memory reference,
with no recorded swapping. The small increases remain relevant, but
these measurements support resident-memory stability over the observed
four-million-frame workloads. PSS excludes other
host memory, so it does not establish a bound on total system RAM use.

Figure~\ref{fig:exp3-cpu} shows mean process CPU occupation of approximately
one logical core over the extended active phases. Configuration
means are 107.03\% at 600~MHz, 105.55\% at 1000~MHz, and
100.21\% at 1500~MHz. The normalization permits values above 100\%
when execution spans multiple threads; it does not imply a single-core
affinity constraint. The lower panels show appreciable activity on all
four cores. This host-wide activity cannot be attributed solely to
\tool{}, and the resource-monitoring workload is not independently
subtracted. The results establish the observed process CPU cost of
sustained acquisition, without treating one-core normalization as a
measurement of total host load or energy consumption.

\subsubsection{Temperature Rise During Hour-Scale Operation}

For each trial, SoC temperature rise is referenced to the sample mean
of its pre-acquisition baseline. All retained traces use the
\texttt{cpu-thermal} sensor; these are SoC readings rather than
individual-core temperatures. The configuration summaries use the mean
active temperature minus that baseline, with original samples retained
for the maximum absolute temperature.

Figure~\ref{fig:exp3-temperature} documents the thermal response during
the extended acquisition workload at each CPU ceiling. Mean active rises
are $7.75\pm0.41$, $12.14\pm0.19$, and
$15.82\pm0.13$~$^\circ$C at 600, 1000, and 1500~MHz,
respectively, with maximum sampled absolute temperatures of 61.84,
65.24, and 68.65~$^\circ$C. The mean trajectories approach elevated
plateaus. At 600~MHz, the wider SD band after roughly 30~min reflects
increased variation between sessions. Inspection of the original traces
shows that one 600-MHz trial departs markedly from the other nine after roughly
30~min and falls substantially later in the session; the available
measurements do not establish the cause. Configuration means therefore
do not describe every session's thermal trajectory. All sessions still
remain available for verification. The larger mean rise at higher ceilings
is a deployment consideration alongside retention and CPU occupation.
Temperature
measurements alone do not establish electrical power, energy per record,
or the absence of frequency throttling, and the configured cooldown
limits transfer of these results to uninterrupted operation without
equivalent thermal conditions.

\subsubsection{Endurance Interpretation}

Taken together, completion of all thirty hour-scale acquisitions,
99.982086\% aggregate sequence retention, a sampled process footprint
below 81~MiB, and the absence of recorded swapping or collector failure
support \tool{} as a sustained acquisition component on the tested
CM4 platform. This conclusion includes the 600-MHz configuration and
does not depend on discarding the larger-loss sessions. It applies to
the recorded build, driver, radio parameters, and storage path at a
1000-frames/s target. Separate process launches and cooldown intervals,
the single enabled receiver, and the finite per-session duration leave
uninterrupted multi-day capture and other deployment conditions for
future evaluation.


\section{Limitations and Future Work}

\tool{} is currently limited to Intel AX200/AX210 interfaces and
requires a CSI-enabled \texttt{backport-iwlwifi} build compatible
with the host kernel. The evaluation covers CM4 hosts, VHT/80-MHz
traffic, and two spatial streams; other PHY formats, channel widths,
radios, and amd64 performance remain outside its scope. Exact
driver revisions and CSI-enabling changes must accompany future
reproduction packages.

The software crossover is a package-level comparison on fixed
device/driver combinations. It demonstrates a repeated performance
ordering but does not independently vary drivers, hardware, and
userspace. Sequence accounting quantifies missing expected identifiers
without locating losses within the acquisition path. Injector-reported
sent counts and retained verifier logs strengthen the audit, but
raw CSI binaries and independent over-the-air transmission traces
are not included. Verification of active files also leaves possible
pending-write effects unresolved.

Ten consecutive repetitions per condition and an ordered sweep
limit inference about independence, environmental drift, and
thermal history. Unequal burst durations affect temperature
comparisons, even after baseline subtraction. Resource samplers
and SSH control introduce load that is common to the experiment
but is not separately quantified. Process PSS excludes other
system memory; polling CPU occupation and SoC
temperature cannot substitute for measured energy consumption.
Some exported UTC timestamps differ between devices and campaign
metadata; analysis uses each trace's relative elapsed time and
does not infer cross-host event coincidence from absolute clocks.

The FeitCSI stress experiment comprises ten completed 600-MHz runs
and one 1500-MHz run that ends in an acquisition-process crash.
The failure traces are empty, and no kernel
out-of-memory evidence is retained. The completed runs establish the
600-MHz memory-growth behavior, but the failed run does not establish
the 1500-MHz memory trajectory or failure mechanism. Memory pressure is
plausible in view of the repeated near-capacity 600-MHz measurements, but
remains unproven and does not establish a specific allocator defect or a
universal maximum duration. The VesperCSI endurance
result
consists of thirty separate hour-scale sessions with cooldown;
it does not establish uninterrupted multi-day operation. Its
larger-loss trials remain relevant, even when overall retention
is high.

Preallocated pools provide bounded buffering rather than lossless
capture. Their capacities are selected before compilation; a
slow synchronous plugin can still create backpressure, triggering
the configured discard or fail-fast policy. No plugin workload
is exercised here. Further evaluation will vary drivers independently,
measure allocation and queue behavior, retain kernel failure
logs and raw captures, quantify power and storage throughput,
and evaluate longer uninterrupted sessions. We also plan to develop
interfaces for Python and MATLAB, with Lua as a possible additional
target, to integrate \tool{} with analysis and experiment-control
workflows. These language interfaces
could broaden the system's use in edge-hosted sensing.

\FloatBarrier

\section{Conclusion}

This paper presented \tool{}, an edge-focused CSI acquisition
system for Intel AX200/AX210 interfaces. Its userspace pipeline
combines fixed-capacity preallocated pools, record validation,
explicit overflow handling, a managed capture lifecycle, and a
versioned plugin interface. Together, these mechanisms provide a
bounded-buffer foundation for separating CSI reception from
experiment-specific consumption.

The evaluations show both the performance opportunity and the
operational trade-offs. In the 600-MHz, 2500-frames/s burst
condition, \tool{} retained 99.017\% and 99.087\% of the expected
10,000-frame sequences, compared with 46.553\% and 43.972\% for
FeitCSI; this ordering persisted when the userspace programs were
exchanged between receivers while their drivers remained fixed.
The comparison is therefore evidence about the tested software-and-
driver packages, not an isolated userspace effect. In a separate
FeitCSI stress study, the 600-MHz trials showed substantial
resident-plus-swapped memory growth, while the higher-ceiling trial
ended in an acquisition-process crash whose cause was not
established.

Across thirty separate hour-scale \tool{} sessions, each injecting
four million frames at 1000~frames/s, aggregate sequence retention
was 99.9821\%. Sampled process PSS remained at or below
80.792~MiB, with no recorded swapping or acquisition-process
failure. These results demonstrate high-retention, extended CSI
acquisition on the evaluated Raspberry Pi CM4 platform. They do
not establish lossless capture or performance across other hosts,
drivers, workloads, or uninterrupted multi-day operation; the
observed CPU and thermal costs, occasional losses, and study scope
remain important deployment constraints.

\FloatBarrier
\bibliographystyle{IEEEtran}
\bibliography{references}

\end{document}

%% file: tables/experiment1_packet_loss.tex
\begin{table*}[t]
\centering
\caption{Experiment 1: missing expected sequence identifiers per 10,000-frame trial. Entries give mean $\pm$ sample standard deviation over ten repetitions. Assignments are reported separately. CPU values are all-core frequency ceilings.}
\label{tab:exp1-loss}
\scriptsize
\setlength{\tabcolsep}{1pt}
\begin{tabular}{@{}rrcccc@{}}
\toprule
 & & \multicolumn{2}{c}{Assignment I} & \multicolumn{2}{c}{Assignment II} \\
\cmidrule(lr){3-4}\cmidrule(lr){5-6}
CPU (MHz) & Target (Hz) & VesperCSI on A & FeitCSI on B & VesperCSI on B & FeitCSI on A \\
\midrule
600 & 100 & $18.8 \pm 7.1$ & $47.8 \pm 13.6$ & $23.5 \pm 9.8$ & $75.0 \pm 17.7$ \\
600 & 200 & $13.4 \pm 5.5$ & $258.7 \pm 9.8$ & $10.7 \pm 3.6$ & $376.2 \pm 23.1$ \\
600 & 500 & $7.7 \pm 4.0$ & $837.7 \pm 99.4$ & $6.1 \pm 2.1$ & $837.3 \pm 38.3$ \\
600 & 1000 & $9.2 \pm 4.2$ & $1328.1 \pm 87.1$ & $7.7 \pm 3.5$ & $1529.0 \pm 23.5$ \\
600 & 1500 & $28.5 \pm 3.8$ & $2399.6 \pm 183.9$ & $18.7 \pm 4.9$ & $2647.2 \pm 104.1$ \\
600 & 2000 & $57.0 \pm 12.3$ & $4118.4 \pm 112.2$ & $46.7 \pm 7.8$ & $4483.6 \pm 120.5$ \\
600 & 2500 & $98.3 \pm 13.2$ & $5344.7 \pm 180.8$ & $91.3 \pm 11.7$ & $5602.8 \pm 66.4$ \\
\midrule
800 & 100 & $21.5 \pm 11.5$ & $37.2 \pm 18.6$ & $12.6 \pm 3.7$ & $43.4 \pm 34.2$ \\
800 & 200 & $12.5 \pm 5.6$ & $185.8 \pm 33.7$ & $8.8 \pm 4.7$ & $250.5 \pm 30.3$ \\
800 & 500 & $9.2 \pm 7.3$ & $636.0 \pm 49.3$ & $4.7 \pm 2.6$ & $655.5 \pm 55.2$ \\
800 & 1000 & $8.8 \pm 3.1$ & $940.3 \pm 88.2$ & $6.8 \pm 1.0$ & $923.3 \pm 65.9$ \\
800 & 1500 & $25.7 \pm 7.5$ & $1616.9 \pm 118.7$ & $23.6 \pm 8.7$ & $1538.0 \pm 74.2$ \\
800 & 2000 & $57.1 \pm 9.5$ & $2603.1 \pm 338.4$ & $49.7 \pm 8.4$ & $2634.8 \pm 210.2$ \\
800 & 2500 & $114.8 \pm 21.8$ & $3786.1 \pm 160.3$ & $98.2 \pm 11.4$ & $3984.6 \pm 121.3$ \\
\midrule
1000 & 100 & $18.9 \pm 5.8$ & $34.8 \pm 11.3$ & $14.3 \pm 4.0$ & $39.9 \pm 45.8$ \\
1000 & 200 & $11.7 \pm 3.6$ & $161.7 \pm 30.9$ & $7.7 \pm 3.2$ & $195.9 \pm 58.1$ \\
1000 & 500 & $11.4 \pm 6.4$ & $566.6 \pm 43.8$ & $6.2 \pm 4.6$ & $539.7 \pm 41.8$ \\
1000 & 1000 & $8.2 \pm 2.9$ & $827.2 \pm 60.6$ & $7.2 \pm 3.4$ & $862.8 \pm 319.7$ \\
1000 & 1500 & $22.3 \pm 6.0$ & $1258.0 \pm 166.9$ & $22.3 \pm 3.9$ & $1245.3 \pm 91.7$ \\
1000 & 2000 & $48.4 \pm 6.3$ & $1845.8 \pm 280.4$ & $50.8 \pm 8.8$ & $1800.8 \pm 215.9$ \\
1000 & 2500 & $86.0 \pm 11.6$ & $2533.1 \pm 438.3$ & $99.9 \pm 10.0$ & $2617.2 \pm 197.9$ \\
\midrule
1200 & 100 & $17.8 \pm 3.7$ & $36.4 \pm 22.8$ & $13.8 \pm 3.3$ & $23.1 \pm 7.7$ \\
1200 & 200 & $12.4 \pm 4.9$ & $129.2 \pm 24.0$ & $7.9 \pm 4.6$ & $148.9 \pm 47.3$ \\
1200 & 500 & $8.7 \pm 7.9$ & $482.6 \pm 40.6$ & $5.0 \pm 2.4$ & $447.9 \pm 20.1$ \\
1200 & 1000 & $8.2 \pm 2.0$ & $748.9 \pm 37.9$ & $5.8 \pm 3.6$ & $690.9 \pm 123.0$ \\
1200 & 1500 & $24.5 \pm 5.6$ & $1001.9 \pm 47.7$ & $22.0 \pm 4.2$ & $984.4 \pm 45.3$ \\
1200 & 2000 & $52.8 \pm 9.6$ & $1414.9 \pm 191.8$ & $49.5 \pm 7.0$ & $1334.8 \pm 134.3$ \\
1200 & 2500 & $116.6 \pm 36.9$ & $1984.3 \pm 281.7$ & $96.3 \pm 12.2$ & $2081.2 \pm 611.3$ \\
\midrule
1500 & 100 & $19.1 \pm 8.2$ & $31.3 \pm 25.9$ & $13.6 \pm 3.7$ & $44.1 \pm 48.5$ \\
1500 & 200 & $11.7 \pm 3.2$ & $116.5 \pm 28.9$ & $7.0 \pm 2.4$ & $89.0 \pm 11.2$ \\
1500 & 500 & $8.0 \pm 5.0$ & $395.6 \pm 34.4$ & $6.9 \pm 2.6$ & $393.3 \pm 48.7$ \\
1500 & 1000 & $8.0 \pm 3.0$ & $613.4 \pm 42.9$ & $4.6 \pm 2.0$ & $579.7 \pm 38.8$ \\
1500 & 1500 & $23.9 \pm 10.5$ & $834.3 \pm 35.6$ & $19.0 \pm 2.4$ & $778.2 \pm 69.7$ \\
1500 & 2000 & $61.4 \pm 16.5$ & $1146.4 \pm 83.1$ & $52.3 \pm 6.7$ & $1326.8 \pm 706.6$ \\
1500 & 2500 & $108.0 \pm 26.8$ & $1711.3 \pm 217.6$ & $98.4 \pm 19.4$ & $1619.9 \pm 262.7$ \\
\bottomrule
\end{tabular}
\end{table*}

%% file: tables/experiment1_resources.tex
\begin{table*}[t]
\centering
\caption{Experiment 1 resource summaries, assignment I. Each CPU/target-rate/tool entry gives mean $\pm$ sample standard deviation across ten repetitions of that configuration. $C$: trial mean process CPU percentage normalized to one logical core; $M$: trial sample mean of PSS + SwapPSS in MiB, summed at each sampling instant; $\Delta T$: trial mean active SoC temperature minus its mean baseline in $^\circ$C. All measures span the controller's active phase, including startup, injection, drain, verification, and termination. SDs describe between-trial variation within each configuration, not standard errors or confidence intervals.}
\label{tab:exp1-resources}
\scriptsize
\setlength{\tabcolsep}{1pt}
\begin{tabular}{@{}rrrrrrrr@{}}
\toprule
 & & \multicolumn{3}{c}{VesperCSI on A} & \multicolumn{3}{c}{FeitCSI on B} \\
\cmidrule(lr){3-5}\cmidrule(lr){6-8}
CPU (MHz) & Target (Hz) & $C$ (\%) & $M$ (MiB) & $\Delta T$ ($^\circ$C) & $C$ (\%) & $M$ (MiB) & $\Delta T$ ($^\circ$C) \\
\midrule
600 & 100 & $100.04 \pm 0.20$ & $78.70 \pm 0.13$ & $2.03 \pm 0.13$ & $8.06 \pm 0.08$ & $216.92 \pm 0.48$ & $1.08 \pm 0.26$ \\
600 & 200 & $100.59 \pm 0.10$ & $78.68 \pm 0.05$ & $1.84 \pm 0.09$ & $12.26 \pm 0.04$ & $220.10 \pm 0.53$ & $0.91 \pm 0.06$ \\
600 & 500 & $100.54 \pm 0.31$ & $78.76 \pm 0.30$ & $1.73 \pm 0.06$ & $17.33 \pm 0.31$ & $221.81 \pm 1.88$ & $0.95 \pm 0.06$ \\
600 & 1000 & $100.42 \pm 0.18$ & $78.73 \pm 0.03$ & $1.64 \pm 0.04$ & $19.53 \pm 0.27$ & $220.58 \pm 2.43$ & $0.97 \pm 0.08$ \\
600 & 1500 & $99.56 \pm 0.64$ & $79.35 \pm 1.20$ & $1.63 \pm 0.07$ & $18.10 \pm 0.39$ & $200.20 \pm 4.50$ & $0.99 \pm 0.06$ \\
600 & 2000 & $99.30 \pm 1.16$ & $80.38 \pm 4.88$ & $1.64 \pm 0.06$ & $14.84 \pm 0.26$ & $161.02 \pm 3.15$ & $0.93 \pm 0.07$ \\
600 & 2500 & $99.52 \pm 0.21$ & $78.90 \pm 0.04$ & $1.61 \pm 0.06$ & $12.39 \pm 0.38$ & $131.80 \pm 4.29$ & $0.99 \pm 0.06$ \\
\midrule
800 & 100 & $100.27 \pm 0.10$ & $78.65 \pm 0.04$ & $2.54 \pm 0.51$ & $6.39 \pm 0.05$ & $214.86 \pm 0.57$ & $1.20 \pm 0.28$ \\
800 & 200 & $100.44 \pm 0.22$ & $78.70 \pm 0.19$ & $2.13 \pm 0.05$ & $9.80 \pm 0.06$ & $218.94 \pm 1.10$ & $1.01 \pm 0.06$ \\
800 & 500 & $100.74 \pm 0.21$ & $78.68 \pm 0.05$ & $2.07 \pm 0.08$ & $14.34 \pm 0.22$ & $223.50 \pm 1.41$ & $1.07 \pm 0.05$ \\
800 & 1000 & $100.42 \pm 0.33$ & $78.80 \pm 0.31$ & $2.07 \pm 0.11$ & $16.43 \pm 0.24$ & $226.74 \pm 2.64$ & $1.09 \pm 0.06$ \\
800 & 1500 & $99.94 \pm 0.22$ & $79.08 \pm 0.80$ & $2.05 \pm 0.08$ & $16.24 \pm 0.34$ & $217.07 \pm 2.25$ & $1.10 \pm 0.07$ \\
800 & 2000 & $99.97 \pm 0.29$ & $78.82 \pm 0.12$ & $2.05 \pm 0.08$ & $14.59 \pm 0.49$ & $196.09 \pm 8.53$ & $1.07 \pm 0.06$ \\
800 & 2500 & $99.56 \pm 0.47$ & $78.81 \pm 0.06$ & $2.09 \pm 0.07$ & $12.73 \pm 0.35$ & $168.73 \pm 4.00$ & $1.09 \pm 0.08$ \\
\midrule
1000 & 100 & $100.50 \pm 0.15$ & $78.65 \pm 0.03$ & $2.93 \pm 0.34$ & $5.56 \pm 0.04$ & $213.77 \pm 0.80$ & $1.11 \pm 0.18$ \\
1000 & 200 & $100.76 \pm 0.04$ & $78.64 \pm 0.01$ & $2.60 \pm 0.08$ & $8.49 \pm 0.10$ & $218.29 \pm 0.69$ & $1.03 \pm 0.06$ \\
1000 & 500 & $100.91 \pm 0.17$ & $78.68 \pm 0.10$ & $2.38 \pm 0.07$ & $12.55 \pm 0.17$ & $223.40 \pm 1.57$ & $1.12 \pm 0.06$ \\
1000 & 1000 & $100.70 \pm 0.19$ & $78.71 \pm 0.07$ & $2.31 \pm 0.09$ & $14.63 \pm 0.13$ & $227.98 \pm 1.80$ & $1.11 \pm 0.07$ \\
1000 & 1500 & $100.38 \pm 0.17$ & $78.72 \pm 0.04$ & $2.31 \pm 0.08$ & $14.70 \pm 0.31$ & $223.57 \pm 3.77$ & $1.15 \pm 0.06$ \\
1000 & 2000 & $100.00 \pm 0.25$ & $78.75 \pm 0.05$ & $2.36 \pm 0.06$ & $14.13 \pm 0.31$ & $212.52 \pm 7.02$ & $1.16 \pm 0.07$ \\
1000 & 2500 & $99.85 \pm 0.19$ & $78.79 \pm 0.03$ & $2.32 \pm 0.07$ & $13.08 \pm 0.66$ & $198.04 \pm 10.40$ & $1.10 \pm 0.10$ \\
\midrule
1200 & 100 & $100.42 \pm 0.09$ & $78.67 \pm 0.05$ & $3.38 \pm 0.35$ & $5.14 \pm 0.05$ & $213.31 \pm 0.74$ & $1.11 \pm 0.13$ \\
1200 & 200 & $100.67 \pm 0.21$ & $78.72 \pm 0.12$ & $2.97 \pm 0.07$ & $7.76 \pm 0.08$ & $217.27 \pm 0.65$ & $1.06 \pm 0.09$ \\
1200 & 500 & $100.99 \pm 0.10$ & $78.65 \pm 0.00$ & $2.75 \pm 0.08$ & $11.44 \pm 0.13$ & $223.31 \pm 0.84$ & $1.13 \pm 0.08$ \\
1200 & 1000 & $100.40 \pm 1.26$ & $78.67 \pm 0.01$ & $2.64 \pm 0.10$ & $13.42 \pm 0.17$ & $227.98 \pm 1.71$ & $1.19 \pm 0.08$ \\
1200 & 1500 & $100.22 \pm 0.62$ & $79.30 \pm 1.93$ & $2.60 \pm 0.10$ & $13.81 \pm 0.13$ & $226.87 \pm 1.53$ & $1.23 \pm 0.06$ \\
1200 & 2000 & $100.09 \pm 0.22$ & $78.72 \pm 0.02$ & $2.60 \pm 0.11$ & $13.67 \pm 0.13$ & $220.78 \pm 4.40$ & $1.23 \pm 0.10$ \\
1200 & 2500 & $99.77 \pm 0.93$ & $80.07 \pm 4.17$ & $2.69 \pm 0.10$ & $12.96 \pm 0.30$ & $208.47 \pm 6.68$ & $1.24 \pm 0.07$ \\
\midrule
1500 & 100 & $100.26 \pm 0.29$ & $78.72 \pm 0.11$ & $4.13 \pm 0.35$ & $4.68 \pm 0.13$ & $213.28 \pm 0.80$ & $1.18 \pm 0.11$ \\
1500 & 200 & $100.43 \pm 0.27$ & $78.82 \pm 0.31$ & $3.71 \pm 0.13$ & $6.96 \pm 0.10$ & $217.28 \pm 0.74$ & $1.24 \pm 0.05$ \\
1500 & 500 & $100.44 \pm 0.85$ & $78.65 \pm 0.02$ & $3.34 \pm 0.13$ & $10.08 \pm 0.18$ & $225.19 \pm 1.06$ & $1.28 \pm 0.07$ \\
1500 & 1000 & $99.79 \pm 0.95$ & $78.67 \pm 0.02$ & $3.19 \pm 0.13$ & $11.89 \pm 0.09$ & $230.98 \pm 1.33$ & $1.31 \pm 0.11$ \\
1500 & 1500 & $99.49 \pm 0.57$ & $78.68 \pm 0.02$ & $3.13 \pm 0.10$ & $12.27 \pm 0.10$ & $230.97 \pm 1.24$ & $1.33 \pm 0.06$ \\
1500 & 2000 & $99.51 \pm 0.47$ & $78.71 \pm 0.02$ & $3.14 \pm 0.11$ & $12.09 \pm 0.18$ & $226.30 \pm 2.33$ & $1.30 \pm 0.08$ \\
1500 & 2500 & $99.22 \pm 0.57$ & $79.32 \pm 1.48$ & $3.15 \pm 0.09$ & $11.63 \pm 0.23$ & $215.55 \pm 4.93$ & $1.32 \pm 0.09$ \\
\bottomrule
\end{tabular}
\end{table*}

\begin{table*}[t]
\centering
\caption{Experiment 1 resource summaries, assignment II. Each CPU/target-rate/tool entry gives mean $\pm$ sample standard deviation across ten repetitions of that configuration. $C$: trial mean process CPU percentage normalized to one logical core; $M$: trial sample mean of PSS + SwapPSS in MiB, summed at each sampling instant; $\Delta T$: trial mean active SoC temperature minus its mean baseline in $^\circ$C. All measures span the controller's active phase, including startup, injection, drain, verification, and termination. SDs describe between-trial variation within each configuration, not standard errors or confidence intervals.}
\label{tab:exp1-resources-II}
\scriptsize
\setlength{\tabcolsep}{1pt}
\begin{tabular}{@{}rrrrrrrr@{}}
\toprule
 & & \multicolumn{3}{c}{VesperCSI on B} & \multicolumn{3}{c}{FeitCSI on A} \\
\cmidrule(lr){3-5}\cmidrule(lr){6-8}
CPU (MHz) & Target (Hz) & $C$ (\%) & $M$ (MiB) & $\Delta T$ ($^\circ$C) & $C$ (\%) & $M$ (MiB) & $\Delta T$ ($^\circ$C) \\
\midrule
600 & 100 & $99.53 \pm 0.10$ & $78.65 \pm 0.04$ & $2.04 \pm 0.41$ & $8.57 \pm 0.05$ & $208.78 \pm 0.43$ & $1.10 \pm 0.25$ \\
600 & 200 & $99.75 \pm 0.24$ & $78.78 \pm 0.21$ & $1.68 \pm 0.06$ & $12.68 \pm 0.10$ & $204.92 \pm 1.26$ & $0.98 \pm 0.09$ \\
600 & 500 & $99.79 \pm 0.18$ & $78.70 \pm 0.07$ & $1.68 \pm 0.11$ & $17.56 \pm 0.27$ & $199.20 \pm 1.67$ & $1.03 \pm 0.05$ \\
600 & 1000 & $99.55 \pm 0.42$ & $79.24 \pm 1.10$ & $1.67 \pm 0.06$ & $18.93 \pm 0.32$ & $190.93 \pm 0.48$ & $1.01 \pm 0.08$ \\
600 & 1500 & $99.41 \pm 0.29$ & $78.83 \pm 0.16$ & $1.64 \pm 0.08$ & $17.14 \pm 0.25$ & $170.21 \pm 2.29$ & $1.07 \pm 0.08$ \\
600 & 2000 & $99.18 \pm 0.41$ & $78.82 \pm 0.05$ & $1.65 \pm 0.07$ & $13.81 \pm 0.25$ & $132.91 \pm 3.38$ & $1.09 \pm 0.08$ \\
600 & 2500 & $98.86 \pm 0.29$ & $78.93 \pm 0.08$ & $1.72 \pm 0.04$ & $11.51 \pm 0.26$ & $110.22 \pm 2.20$ & $1.04 \pm 0.05$ \\
\midrule
800 & 100 & $99.78 \pm 0.10$ & $78.64 \pm 0.03$ & $2.51 \pm 0.47$ & $6.67 \pm 0.07$ & $209.68 \pm 0.97$ & $1.26 \pm 0.44$ \\
800 & 200 & $100.02 \pm 0.15$ & $78.69 \pm 0.05$ & $2.19 \pm 0.13$ & $10.03 \pm 0.06$ & $206.78 \pm 1.08$ & $1.11 \pm 0.08$ \\
800 & 500 & $100.01 \pm 0.39$ & $78.67 \pm 0.01$ & $1.97 \pm 0.05$ & $14.32 \pm 0.27$ & $202.60 \pm 1.66$ & $1.15 \pm 0.06$ \\
800 & 1000 & $99.62 \pm 0.51$ & $79.20 \pm 1.46$ & $1.93 \pm 0.04$ & $16.10 \pm 0.30$ & $201.66 \pm 1.97$ & $1.16 \pm 0.08$ \\
800 & 1500 & $99.50 \pm 0.33$ & $78.80 \pm 0.22$ & $1.87 \pm 0.05$ & $15.82 \pm 0.19$ & $191.78 \pm 1.58$ & $1.15 \pm 0.08$ \\
800 & 2000 & $99.54 \pm 0.27$ & $78.77 \pm 0.04$ & $1.90 \pm 0.10$ & $14.04 \pm 0.39$ & $170.35 \pm 4.45$ & $1.13 \pm 0.06$ \\
800 & 2500 & $99.24 \pm 0.53$ & $79.44 \pm 1.98$ & $1.95 \pm 0.05$ & $11.95 \pm 0.30$ & $142.54 \pm 2.04$ & $1.18 \pm 0.07$ \\
\midrule
1000 & 100 & $99.96 \pm 0.19$ & $78.68 \pm 0.07$ & $2.83 \pm 0.34$ & $5.67 \pm 0.07$ & $209.22 \pm 0.62$ & $1.24 \pm 0.23$ \\
1000 & 200 & $100.17 \pm 0.19$ & $78.70 \pm 0.10$ & $2.47 \pm 0.10$ & $8.57 \pm 0.08$ & $207.54 \pm 0.94$ & $1.14 \pm 0.10$ \\
1000 & 500 & $100.29 \pm 0.27$ & $78.65 \pm 0.01$ & $2.31 \pm 0.06$ & $12.48 \pm 0.27$ & $206.03 \pm 1.48$ & $1.16 \pm 0.11$ \\
1000 & 1000 & $100.00 \pm 0.40$ & $78.68 \pm 0.01$ & $2.23 \pm 0.10$ & $14.08 \pm 0.50$ & $203.07 \pm 6.34$ & $1.21 \pm 0.07$ \\
1000 & 1500 & $99.87 \pm 0.25$ & $78.94 \pm 0.69$ & $2.20 \pm 0.08$ & $14.20 \pm 0.18$ & $197.11 \pm 2.41$ & $1.22 \pm 0.08$ \\
1000 & 2000 & $99.61 \pm 0.22$ & $78.75 \pm 0.03$ & $2.19 \pm 0.09$ & $13.62 \pm 0.26$ & $187.11 \pm 4.77$ & $1.20 \pm 0.08$ \\
1000 & 2500 & $99.42 \pm 0.24$ & $78.99 \pm 0.63$ & $2.17 \pm 0.06$ & $12.31 \pm 0.33$ & $171.29 \pm 4.32$ & $1.26 \pm 0.07$ \\
\midrule
1200 & 100 & $100.05 \pm 0.23$ & $78.68 \pm 0.11$ & $3.17 \pm 0.33$ & $5.15 \pm 0.07$ & $210.86 \pm 0.63$ & $1.31 \pm 0.17$ \\
1200 & 200 & $100.31 \pm 0.14$ & $78.65 \pm 0.03$ & $2.84 \pm 0.06$ & $7.74 \pm 0.08$ & $209.21 \pm 1.12$ & $1.30 \pm 0.08$ \\
1200 & 500 & $100.36 \pm 0.30$ & $78.65 \pm 0.01$ & $2.63 \pm 0.09$ & $11.24 \pm 0.07$ & $209.14 \pm 1.01$ & $1.30 \pm 0.08$ \\
1200 & 1000 & $99.85 \pm 0.68$ & $79.45 \pm 1.36$ & $2.51 \pm 0.11$ & $13.19 \pm 0.23$ & $210.07 \pm 2.92$ & $1.27 \pm 0.08$ \\
1200 & 1500 & $99.98 \pm 0.31$ & $78.70 \pm 0.03$ & $2.52 \pm 0.11$ & $13.35 \pm 0.19$ & $205.07 \pm 2.08$ & $1.31 \pm 0.12$ \\
1200 & 2000 & $99.42 \pm 0.50$ & $80.13 \pm 2.25$ & $2.52 \pm 0.07$ & $13.22 \pm 0.12$ & $199.15 \pm 2.98$ & $1.34 \pm 0.10$ \\
1200 & 2500 & $99.73 \pm 0.30$ & $78.78 \pm 0.04$ & $2.56 \pm 0.10$ & $12.18 \pm 0.75$ & $184.60 \pm 12.51$ & $1.36 \pm 0.08$ \\
\midrule
1500 & 100 & $99.86 \pm 0.54$ & $78.63 \pm 0.02$ & $3.84 \pm 0.42$ & $4.51 \pm 0.12$ & $210.70 \pm 1.02$ & $1.44 \pm 0.21$ \\
1500 & 200 & $100.36 \pm 0.47$ & $78.72 \pm 0.11$ & $3.38 \pm 0.15$ & $6.71 \pm 0.08$ & $210.66 \pm 0.58$ & $1.37 \pm 0.07$ \\
1500 & 500 & $99.80 \pm 1.29$ & $78.64 \pm 0.01$ & $3.14 \pm 0.16$ & $9.79 \pm 0.23$ & $211.11 \pm 2.46$ & $1.51 \pm 0.08$ \\
1500 & 1000 & $98.99 \pm 0.97$ & $78.67 \pm 0.02$ & $2.99 \pm 0.09$ & $11.43 \pm 0.14$ & $213.09 \pm 1.33$ & $1.45 \pm 0.08$ \\
1500 & 1500 & $99.14 \pm 0.65$ & $78.73 \pm 0.05$ & $2.99 \pm 0.14$ & $11.80 \pm 0.12$ & $212.00 \pm 1.57$ & $1.48 \pm 0.08$ \\
1500 & 2000 & $99.58 \pm 0.72$ & $79.28 \pm 1.74$ & $3.02 \pm 0.08$ & $11.35 \pm 0.87$ & $201.13 \pm 14.37$ & $1.45 \pm 0.07$ \\
1500 & 2500 & $99.40 \pm 0.27$ & $78.74 \pm 0.05$ & $3.05 \pm 0.08$ & $11.19 \pm 0.26$ & $196.22 \pm 4.90$ & $1.47 \pm 0.13$ \\
\bottomrule
\end{tabular}
\end{table*}

%% file: tables/experiment2_packet_loss.tex
\begin{table}[t]
\centering
\caption{Experiment 2: FeitCSI missing expected sequence identifiers on device B at the 600-MHz all-core CPU ceiling and 1000-frames/s target. Each trial expects 100,000 identifiers. Entries give mean $\pm$ sample standard deviation across the ten completed trials.}
\label{tab:exp2-loss}
\scriptsize
\setlength{\tabcolsep}{0pt}
\begin{tabular*}{\columnwidth}{@{\extracolsep{\fill}}lr@{}}
\toprule
CPU (MHz) & 600 \\
Target (Hz) & 1000 \\
Missing/trial & $45483.6 \pm 458.0$ \\
Loss (\%) & $45.484 \pm 0.458$ \\
\bottomrule
\end{tabular*}
\end{table}

%% file: tables/experiment3_retention.tex
\begin{table*}[t]
\centering
\caption{Experiment 3 sequence retention. Each configuration comprises ten trials of four million expected frames at a 1000-frames/s target. Retention uses all 40 million expected identifiers per configuration. }
\label{tab:exp3-retention}
\small
\setlength{\tabcolsep}{3pt}
\begin{tabular}{@{}rrrrrr@{}}
\toprule
CPU (MHz) & Total missing & Retention (\%) \\
\midrule
600 & 5,819 & 99.985453 \\
1000 & 1,089 & 99.997278 \\
1500 & 14,589 & 99.963528 \\
\bottomrule
\end{tabular}
\end{table*}

%% file: tables/experiment3_resources.tex
\begin{table*}[t]
\centering
\caption{Experiment 3 active-phase resource summaries. Mean CPU occupation, trial-mean PSS, endpoint PSS change, and baseline-referenced mean SoC rise are reported as mean $\pm$ sample SD across ten trials. Maxima use original samples over all ten trials. CPU is normalized to one logical core; all four cores remain available. No SwapPSS is recorded.}
\label{tab:exp3-resources}
\scriptsize
\setlength{\tabcolsep}{1pt}
\begin{tabular}{@{}rrrrrrr@{}}
\toprule
CPU (MHz) & Process CPU (\%) & Mean PSS (MiB) & Max PSS (MiB) & PSS change (MiB) & Mean $\Delta T$ ($^\circ$C) & Max $T$ ($^\circ$C) \\
\midrule
600 & $107.03 \pm 0.19$ & $79.15 \pm 0.12$ & 79.77 & $0.80 \pm 0.17$ & $7.75 \pm 0.41$ & 61.84 \\
1000 & $105.55 \pm 0.14$ & $79.25 \pm 0.41$ & 80.79 & $0.93 \pm 0.50$ & $12.14 \pm 0.19$ & 65.24 \\
1500 & $100.21 \pm 0.14$ & $79.10 \pm 0.08$ & 79.51 & $0.76 \pm 0.10$ & $15.82 \pm 0.13$ & 68.65 \\
\bottomrule
\end{tabular}
\end{table*}